\documentclass[]{aa}  
\usepackage{amsmath}
\usepackage{amssymb}
\usepackage{natbib}
\usepackage{graphicx}
\usepackage{txfonts}
\usepackage[english]{babel}
\usepackage{hyperref}
\usepackage{units}
\newcommand{\modified}[1]{#1}
\begin{document}

\title{The color dependent morphology of the post-AGB star HD161796\thanks{Based on observations made with the William Herschel Telescope operated on the island of La Palma by the Isaac Newton Group in the Spanish Observatorio del Roque de los Muchachos of the Instituto de Astrofísica de Canarias.}}
\author{
M. Min\inst{1,2}
	\and
S.~V. Jeffers\inst{3,2}
	\and
H. Canovas\inst{4}
	\and
M. Rodenhuis\inst{2}
	\and
C.~U. Keller\inst{2}
	\and
L.~B.~F.~M. Waters\inst{1,5}
}

\authorrunning{M. Min et al.}

\offprints{M. Min, \email{M.Min@uva.nl}}

\institute{
Astronomical Institute `Anton Pannekoek', University of Amsterdam, 
PO Box 94249, 1090 GE Amsterdam, The Netherlands
	\and
Leiden Observatory, Leiden University, P.O. Box 9513, 2300 RA Leiden, The Netherlands
	\and
Institut f\"ur Astrophysik, Georg-August-Universit\"at, Friedrich-Hund-Platz 1, D-37077 G\"ottingen, Germany
	\and
Departamento de F\'isica y Astronom\'ia, Facultad de Ciencias, Universidad de Valpara\'iso, Valpara\'iso, Chile
	\and
SRON Netherlands Institute for Space Research, Sorbonnelaan 2, 3584 CA Utrecht, The Netherlands
}
   \date{Last revision: \today}

 
  \abstract
   {Many protoplanetary nebulae show strong asymmetries in their surrounding shell, pointing to asymmetries during the mass loss phase. Questions concerning the origin and the onset of deviations from spherical symmetry are important for our understanding of the evolution of these objects. Here we focus on the circumstellar shell of the post-AGB star HD~161796.}
   {We aim at detecting signatures of an aspherical outflow, as well as to derive the properties of it.}
   {We use the imaging polarimeter ExPo (the extreme polarimeter), a visitor instrument at the William Herschel Telescope, to accurately image the dust shell surrounding HD~161796 in various wavelength filters. Imaging polarimetry allows us to separate the faint, polarized, light from circumstellar material from the bright, unpolarized, light from the central star.}
   {The shell around HD~161796 is highly aspherical. A clear signature of an equatorial density enhancement can be seen. This structure is optically thick at short wavelengths and changes its appearance to optically thin at longer wavelengths. In the classification of the two different appearances of planetary nebulae from HST images it changes from being classified as DUPLEX at short wavelengths to SOLE at longer wavelengths. This strengthens the interpretation that these two appearances are manifestations of the same physical structure. Furthermore, we find that the central star is hotter than often assumed and the relatively high observed reddening is due to circumstellar rather than interstellar extinction.}
   {}


   \maketitle
%

\section{Introduction}

When imaged, post-AGB stars often display a highly nonspherical structure \citep{2000ApJ...528..861U}. It has been, and still is, a matter of debate what is the cause for this asphericity and at which point in the post main sequence evolution the deviations from a spherical wind start. One of the most favored explanations is through interaction of the wind with a binary companion. However, recently \citet{2011ApJ...734...25H} have searched for such companions and found only one out of seven proto-planetary nebula contained a detectable binary companion. Detailed images of evolved stars together with extended modeling is required to get a better handle on the origin of the non-spherical nature of many (proto)-planetary nabula.

The types of assymetries seen in protoplanetary nebulae have been devided into two classes by \citet{2000ApJ...528..861U}; the SOLE (Star-Obvious Low-level-Elongated) and the DUPLEX (DUst-Prominent Longitudinally-EXtended) types. The SOLE type shows one elongated structure which was interpreted as the appearance of an optically thin torus surrounding the star. The DUPLEX type on the other hand shows two distinct blobs at opposite sides of the central star. This was interpreted as the appearance of a bipolar outflow structure. In the DUPLEX type there is also a torus, but it is optically thick and thus self-shielding and not visible in scattered light. Essentially, both types are appearances of the same physical structure, where the main difference lies in the optical depth through the torus.

Here we present polarimetric images of the post-AGB object HD~161796 observed with the extreme polarimeter (ExPo) at the William Herschel Telescope (WHT). HD~161796 is an F3Ib star \citep{1984ApJ...285..698F} with a detached circumstellar shell as derived from its double peaked spectral energy distribution (SED) \citep{1989ApJ...346..265H}. The circumstellar matter is clearly observed to be oxygen rich and dominated by silicate emission \citep{1992ApJ...392L..75J, 2002A&A...389..547H}. Spatially resolved images in the near-IR \citep{2000ApJ...528..861U, 2001MNRAS.322..321G}, and the mid-IR \citep{1994ApJ...423L.135S, 2003MNRAS.343..880G} resolve the inner boundary of the shell to be much larger than the dust condensation radius. From these observations it can be derived that the mass loss stopped around 430 years ago.

The shell surrounding HD~161796 has been modeled from different perspectives. \citet{2002A&A...389..547H} provide an extensive analysis of the mineralogy and size distribution of the dust grains in the envelope using a spherical symmetric radiative transfer model. They derive from their fit that the mass loss rate, which ended 430 years ago, was realtively high ($5\cdot10^{-4}M_{\sun}/$yr) which they suggest is consistent with the prominent presence of ice-coated grains. Deviating from spherical symmetry, \citet{2002ApJ...571..936M} developed a general, axisymmetric parameterization to fit images of protoplanetary nebulae. They successfully apply this to HD~161796 and manage to simultaneously fit the SED and the mid IR images. The axisymmetric model by \citet{2002ApJ...571..936M} is very successful in reproducing the various morphologies observed. However, as was shown by \citet{2003MNRAS.343..880G}, there are also indications of deviation from axisymmetry because in the mid-IR images the two lobes east and west from the central star are not equally bright.

We observed HD~161796 using imaging polarimetry with various filters at optical wavelengths. The observed morphology of the circumstellar shell is observed to change with wavelength (section~\ref{sec:observations}). In this paper we present a model which is consistent with the images at different wavelengths and fits the SED from optical to millimeter wavelengths (section~\ref{sec:model}). The wavelength dependent morphology can be understood as a changing optical depth through the equator of the system. This is used to accurately derive the total optical depth through the torus and constrain the scattering properties of the circumstellar dust grains.

\section{Polarimetric imaging using ExPo}
\label{sec:observations}

Imaging polarimetry takes advantage of the fact that scattered light from circumstellar material is polarized to filter out the bright point spread function (PSF) of the central star. This way we can image the surrounding material down to a contrast of $\sim10^{-4}$ \citep{2008SPIE.7014E.227R}. For our observations, the imaging polarimeter ExPo was mounted at the 4.2\,m William Herschel Telescope (WHT). The raw frames from the observations were carefully reduced using a state-of-the-art data reduction pipeline including detailed recentering of the frames to reduce the effects of seeing. The technical setup and data reduction steps are briefly outlined below. For details we refer to \citet{2011A&A...531A.102C}.

\subsection{Technical description}

ExPo is a regular visitor instrument at the William Herschel Telescope, mounted on the Nasmyth platform \citep{2008SPIE.7014E.227R}.
ExPo is a dual-beam imaging polarimeter, operating at visible wavelengths $500-900\,$nm,
that combines fast modulation with the dual-beam technique to minimize systematic errors.  The design concept of ExPo is that
a Ferroelectric Liquid Crystal (FLC) modulates the light, rotating its polarization plane by
90 degrees. The FLC changes its state every 0.028 seconds (approx 35 HZ), which allows us to reduce the atmospheric (seeing) effects.  A cube beamsplitter divides the light into two beams with orthogonal polarization
states. These two simultaneous images are then projected onto an Electron Multiplying Charge Coupled (EM-CCD) camera.

\subsection{Observations}

The observations of HD~161796 were taken in May 2010. To determine the wavelength dependence of the morphology of HD~161796 observations were also secured using a set of different filters. For both the H$_\alpha$ and the Na filters we took images using a filter at the spectral position of the line and in the continuum next to it. No significant differences were detected between the line and continuum observations, so we combined them to increase the signal to noise ratio. Details of the observations and filters used are shown in Table~\ref{tab:obs}.  To obtain a linear polarization image in each filter we take four sub-exposures with FLC angles of 0$^\circ$, 22.5$^\circ$, 45$^\circ$ and 67.5$^\circ$.  For HD~161796, each FLC angle comprises at least 4000 frames of 0.028s duration.  Additionally, a set of calibration flat-fields is taken at the beginning and end of every night and a set of dark frames is taken at the end of each observation.

\begin{table}[!tbp]
\caption{Details of the observations}
\begin{center}
\begin{tabular}{llc}
\hline
\hline
Date			& Filter \\
\hline
25 May 2010	& Na continuum (0.590$\pm0.001\,\mu$m)	\\
25 May 2010	& Na line (0.580$\pm0.001\,\mu$m)	\\
24 May 2010	& H$_\alpha$ continuum (0.670$\pm0.001\,\mu$m)	\\
24 May 2010	& H$_\alpha$ line (0.660$\pm0.001\,\mu$m)	\\
25 May 2010	& Sloan I (0.76$\pm0.08\,\mu$m)	\\
\hline
\end{tabular}
\end{center}
\label{tab:obs}
\end{table}

\subsection{Data analysis}

The FLC modulates the polarization state of the incoming light by 90$^\circ$ every 0.028s switching between "A" and "B" frames. For each frame ("A" or "B") the light is separated, using a beamsplitter, into two orthogonal beams which are simultaneously imaged on the EM-CCD, a concept that is the basis of a dual-beam polarimeter.  For every pair of observations this results in four different images (i.e. in total at least 8000 for each FLC angle as described above).   This results in four different images per FLC cycle ("A"+"B"): $A_\mathrm{left}, A_\mathrm{right}, B_\mathrm{left}, B_\mathrm{right}$.   Each frame is first flat-field corrected, dark subtracted, and cleaned of cosmic rays.  The four images are combined using the double-difference approach \citep{2001ApJ...553L.189K, 2009ApJ...701..804H}.
\begin{equation}
P'_l=0.5(\Delta A-\Delta B)=0.5((A_\mathrm{left}-A_\mathrm{right})-(B_\mathrm{left}-B_\mathrm{right}))
\end{equation}
\modified{where $P'_l$ is the uncalibrated linearly polarised image. We use here $P'_I$ in stead of Q or U since the rotation of the sky and the imperfect alignment of the instrument cause the images obtained in this way to be neither exactly Stokes Q or Stokes U. These images are calibrated to produce the Stokes Q and U images in the reference frame attached to the target.} The total intensity image is computed as:
\begin{equation}
I=0.5(A_\mathrm{left}+A_\mathrm{right}+B_\mathrm{left}+B_\mathrm{right})
\end{equation}
The image alignment algorithm is based on a cross-correlation algorithm
that aligns each image with a template, with an accuracy of a third
of a pixel. The sky polarization is subtracted from the images by computing the polarization of four different sky regions on the images.  The data analysis is described in more detail by \citet{2011A&A...531A.102C}.

Finally, the reduced images are calibrated using the method of
Rodenhuis et al. (in prep) in order to produce Stokes Q and U images.  The
polarised intensity is defined as:$P_l=\sqrt{Q^2+U^2}$ and the degree
of polarisation as $P=P_l/I$.  The polarisation angle, which defines
the orientation of the polarisation plane is $P_\Theta=0.5 \times
\arctan {U}/{Q}$.  The instrumental design of ExPo includes a
polarisation compensator which reduces the instrumental polarization from $\sim$3\%
to approximately \modified{1\%}.  This is further removed in the data
analysis by assuming that the central star is unpolarized. We stress here that the fact that we remove the polarization of the central resolution element causes any polarization present inside the inner resolution element to be removed. This also implies that the polarized intensity we derive over the entire image has to be taken as a lower limit. This effect is described in \citet{2011arXiv1111.4348M}.

\subsection{Polarization images}

\begin{figure*}[!tb]
\centerline{\resizebox{\hsize}{!}{\includegraphics{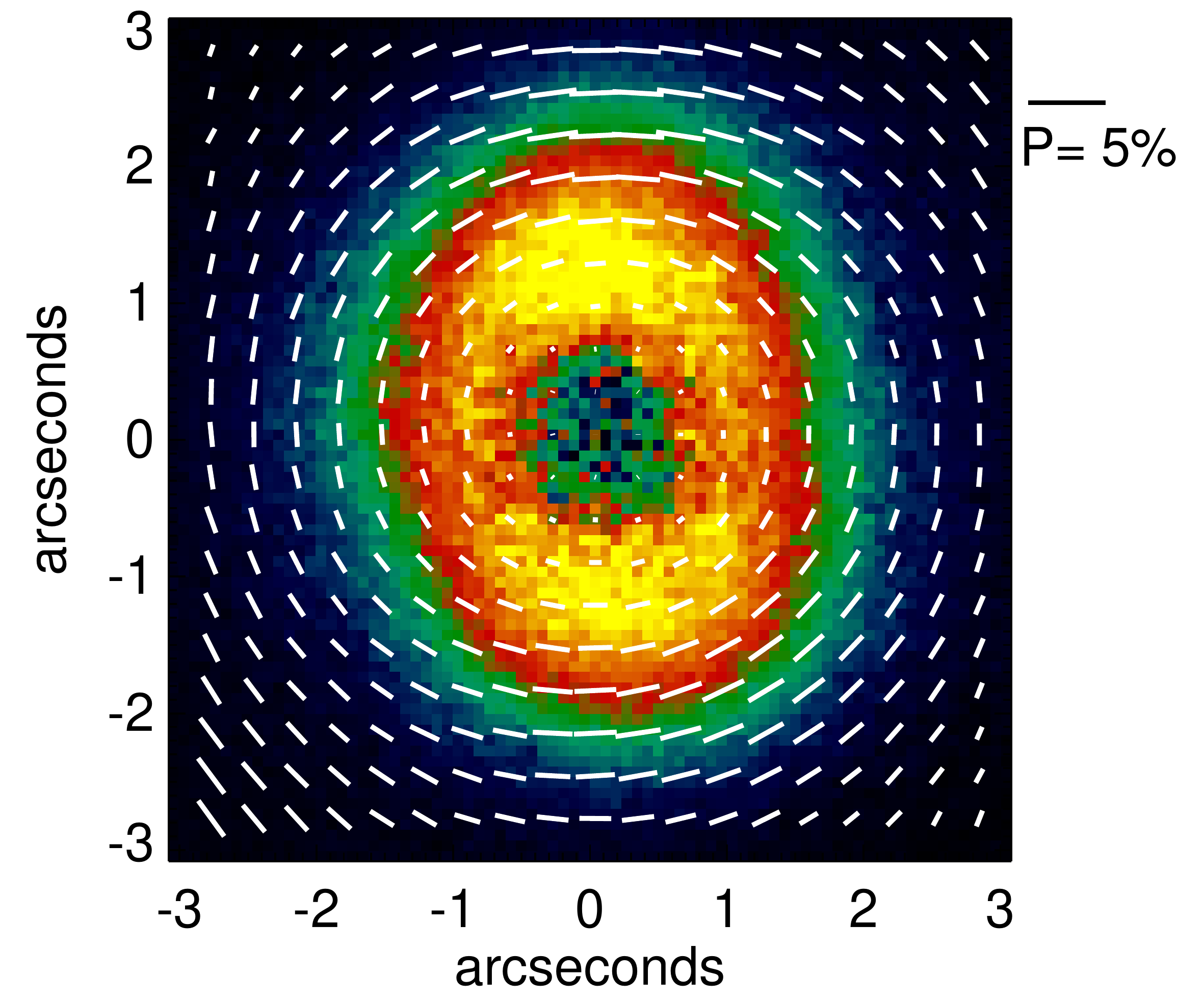}\includegraphics{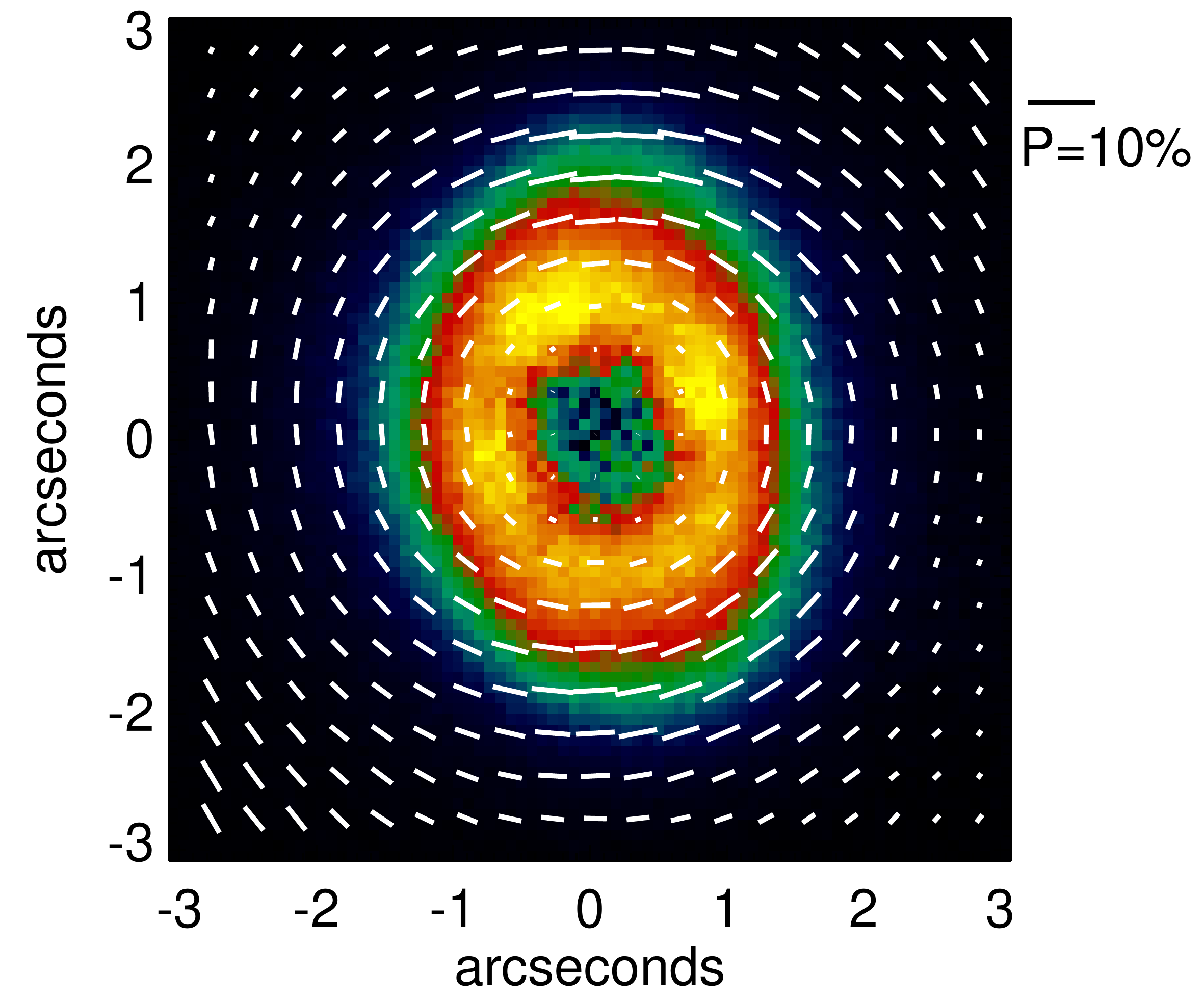}\includegraphics{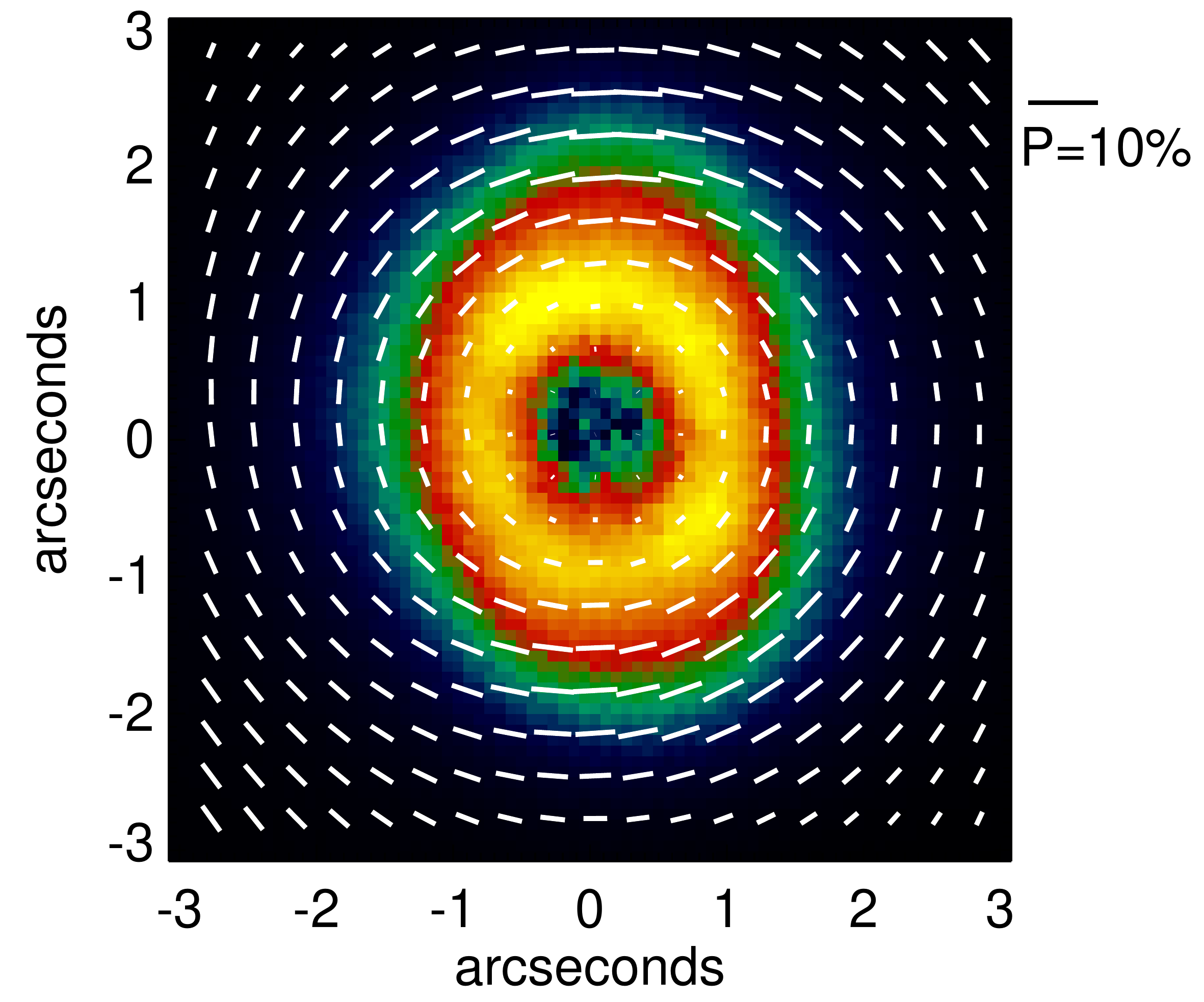}}}
\centerline{\resizebox{\hsize}{!}{\includegraphics{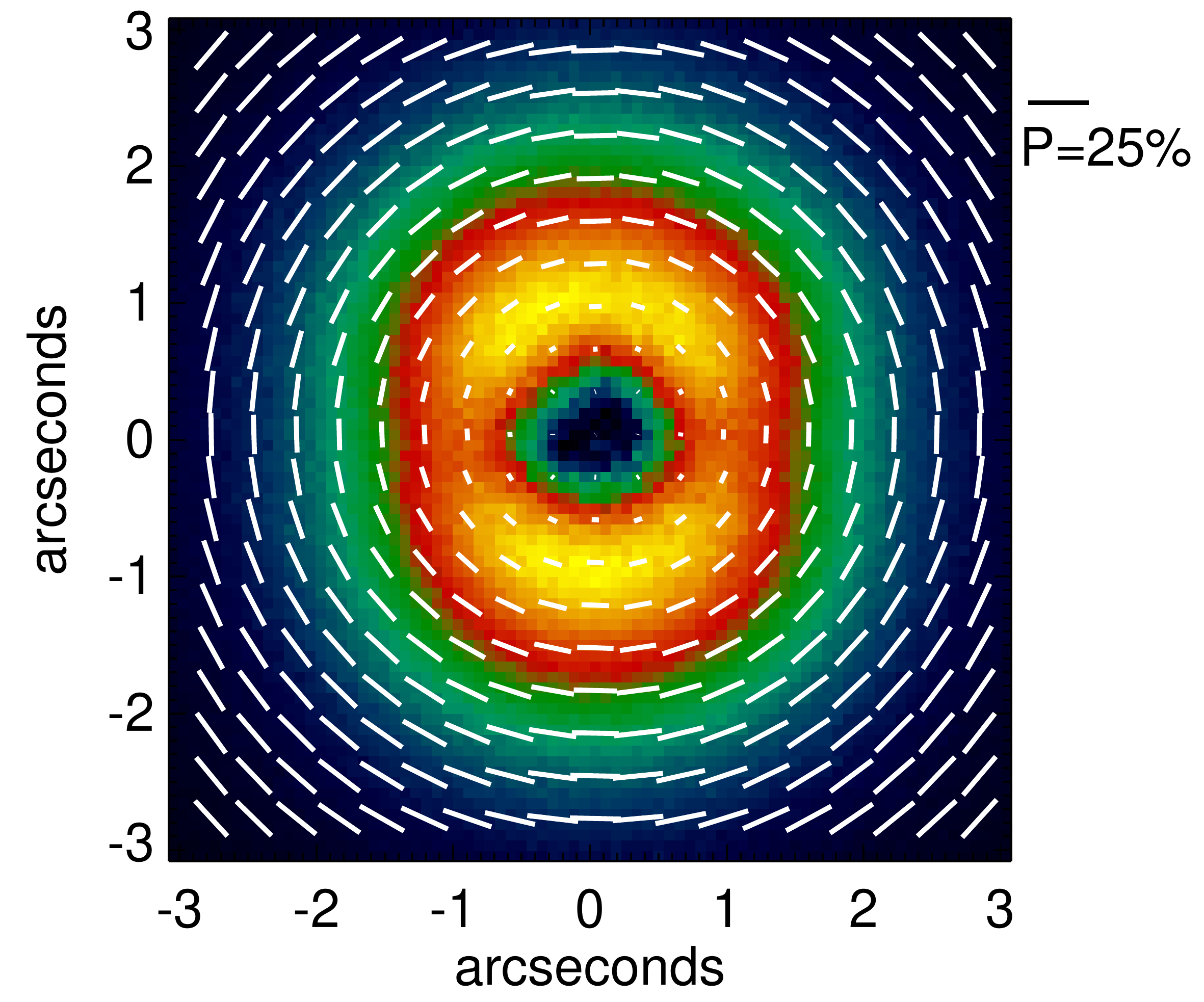}\includegraphics{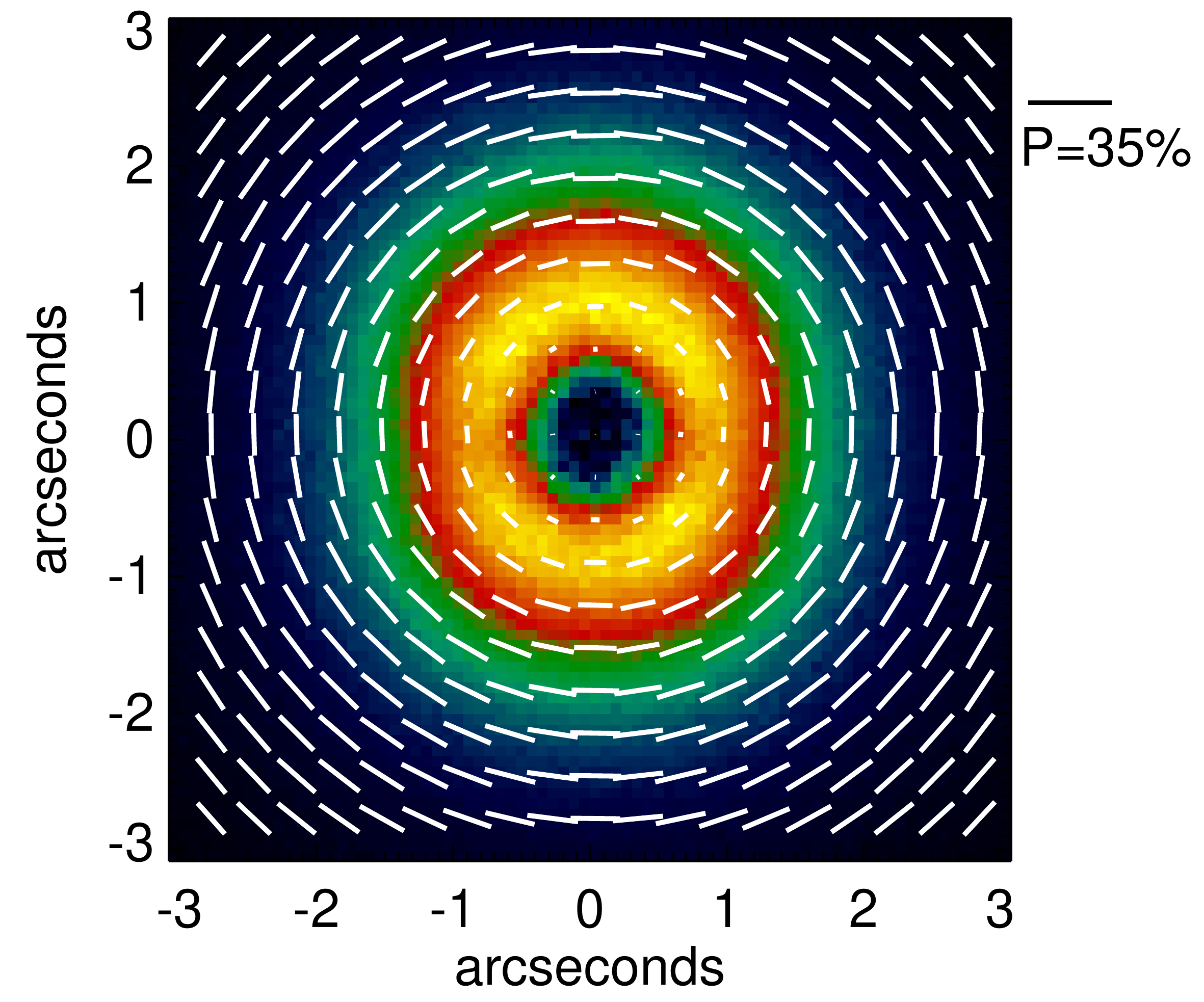}\includegraphics{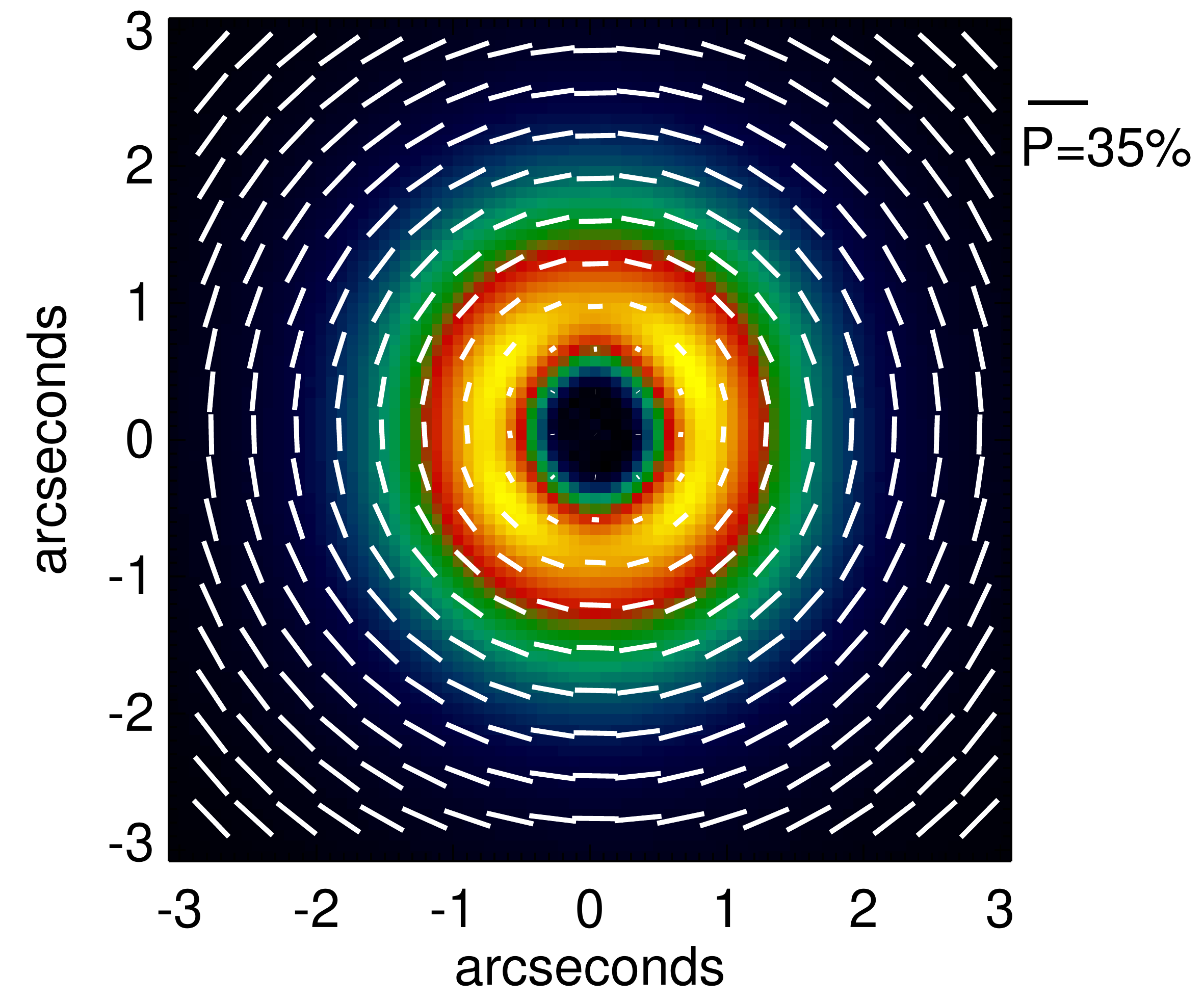}}}
\centerline{\resizebox{7 cm}{!}{\includegraphics{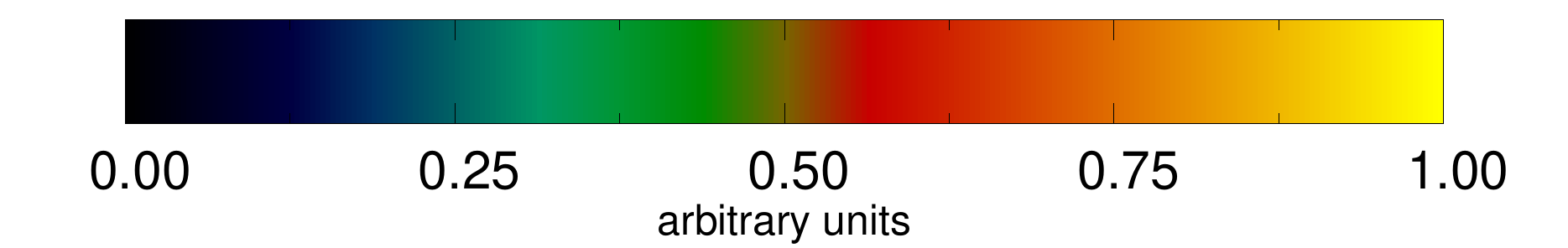}}}
\caption{Polarized intensity images of HD161796 in the different filters (upper panels) as obtained with ExPo. Left panel shows the ExPo observations using the Na 580\,nm filter, middle panel the H$_\alpha$ 656\,nm filter, and the right panels the Sloan I filter. The lower panels display the best fit model images. The absolute scaling of the model images is adjusted to provide a good fit to the observations.}
\label{fig:images}
\end{figure*}

The final polarized intensity images obtained with ExPo for three different filters are displayed in the upper panels of Fig.~\ref{fig:images}. The wavelength increases from left to right. In these images the inner edge of the shell is clearly resolved. \modified{The inner edge of the shell is derived from our images to be at $\sim0.8''$, which translates to 960\,AU at a distance of 1200\,pc \citep{1994ApJ...423L.135S}.} We find a bright ring with an intensity drop which is quite sharp outside of $1.5''$ \modified{(1800\,AU)}. Note that the image taken in the Na filter has a seeing that is significantly worse than the other observations. At the shortest wavelength, the Na filter at 580\,nm, the image displays a brightness enhancement towards the north and south. This structure is consistently seen in both the Na line and the Na continuum filters. The brightness enhancement is not clearly visible in the other two images. The 656\,nm image shows still some structure which is consistently seen in the line and continuum image. \modified{The Southern part of this image seems to be systematically slightly weaker than the Northern part. Also, the Sloan I image still displays some very low level fluctuations in the ring, with a brighter arc to the top and a small brigthness enhancement to the lower right part of the image.} Furthermore, the structure seen in all images is elongated in north-south direction with an aspect ratio of $\sim1.3$.

\modified{The procedure to correct for instrumental polarization alters the inner region of the images. In order to verify that the observed ring structure with an empty inner region is not an effect of this procedure, we have also looked at the images before instrumental polarization correction. We consistently see the ring structure in all images also before the correction. Furthermore, the derived inner edge of the ring, as well as the structure seen in it, is unaffected by the removal of instrumental polarization.}

\section{Modeling the circumstellar dust shell}
\label{sec:model}

In this section we first describe the parameterization chosen for the density setup and the assumptions for the optical properties of the dust grains. Second, we also discuss the radiative transfer model we use.

\subsection{Density setup}

\begin{figure}[!tb]
\centerline{\resizebox{\hsize}{!}{\includegraphics{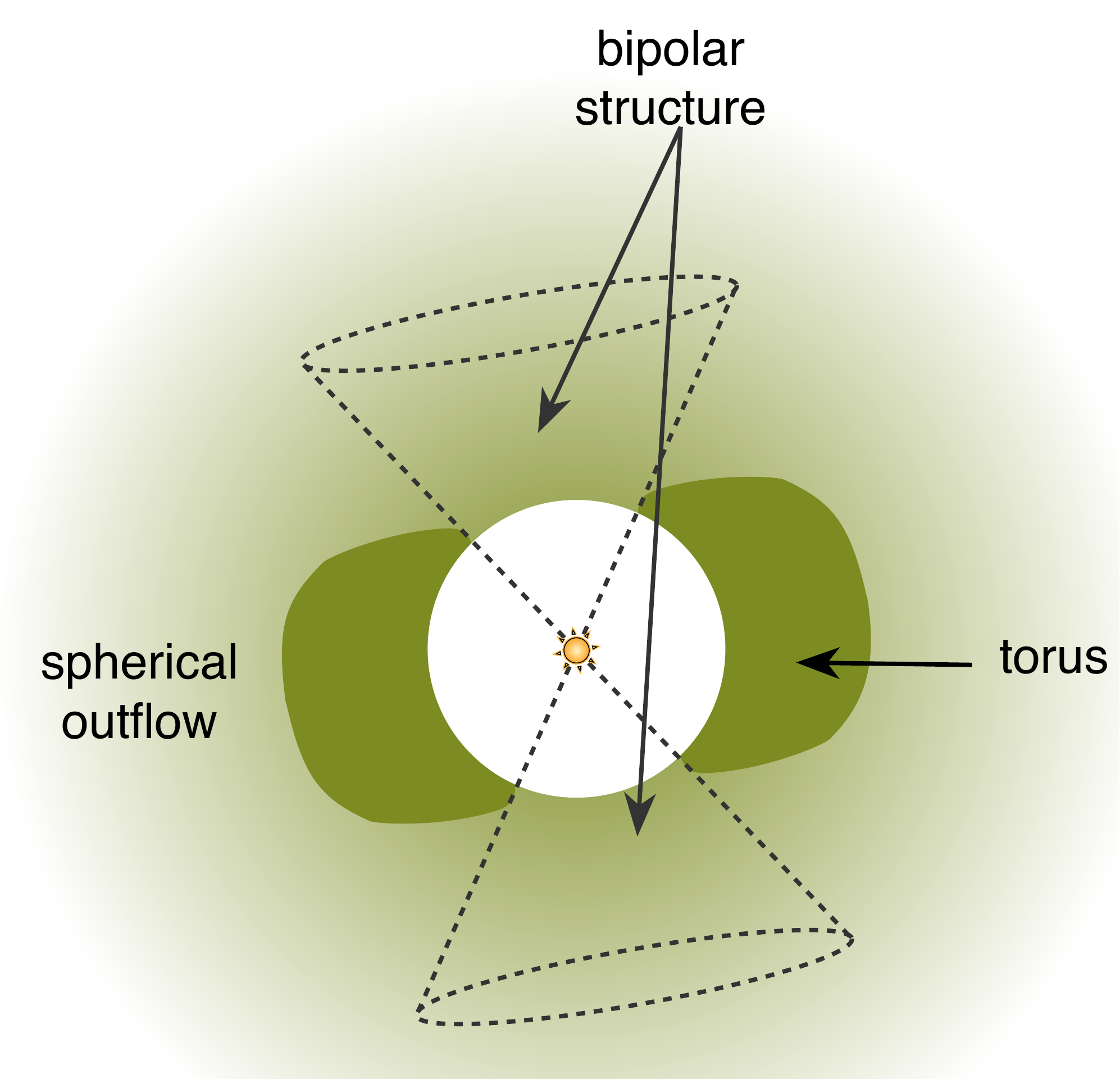}}}
\caption{Cartoon drawing of the density structure around HD~161796. Note that the appearance of the bipolar structure is caused by the fact that the photons escape through the polar regions, illuminating the low density material there.}
\label{fig:cartoon}
\end{figure}

We take the approach from \citet{2002ApJ...571..936M} who find a model which is consistent with the thermal infrared emission (both imaging and the spectral energy distribution). This is a parameterized description of a bipolar structure with an equatorial density enhancement and is given by
\begin{eqnarray}
\rho & = & \rho_0\left(\frac{R}{R_{min}}\right)^{-B\left\{1+C\sin^F\theta\left[\frac{e^{-(R/R_{sw})^D}}{e^{-(R_{min}/R_{sw})^D}}\right]\right\}} \nonumber \\
 && \qquad \times\left\{1+A(1-\cos\theta)^F\left[\frac{e^{-(R/R_{sw})^E}}{e^{-(R_{min}/R_{sw})^E}}\right]\right\},
\end{eqnarray}
were $R$ is the radius to the star, $R_{sw}$ is the radius of the superwind phase, $R_{min}$ is the inner radius of the dust shell, $\theta$ is the angle measured from the pole, $\rho_0$ is the density at the inner edge of the shell, and $A-F$ are fitting parameters. This density setup is empirical, and thus the parameters do not directly relate to physical quantities. The density model from \citet{2002ApJ...571..936M} assumes an equatorial density enhancement caused by the superwind in a more or less torus like structure. In the remaining of this paper we will use the term 'torus' to refer to this equatorial density enhancement. Outside of the superwind phase these authors place a constant mass loss. As discussed below, we adopt their model, and only tune it slightly to account for the differences in dust grain properties we use. A cartoon drawing of the density structure is shown in Fig.~\ref{fig:cartoon}.

\subsection{Dust grain properties}

We have computed the absorption and scattering properties of grains with a mixture of silicates, metallic iron and water ice with mass abundances of 77, 15, and 8\,\% respectively. We do not aim here for an exact compositional analysis, for that we refer to the thorough analysis of the ISO spectrum by \citet{2002A&A...389..547H}. For the composition of the silicates we take the mixture derived by \citet{2007A&A...462..667M} for interstellar silicates. We use the refractive indices for the materials from \citet{1995A&A...300..503D} and \citet{1996A&A...311..291H} for the silicates, from \citet{1988AplO...27.1203O} for the metallic iron, and from \citet{1984ApOpt..23.1206W} for the water ice. The exact grain composition has little influence on the scattering properties of the particles compared to the particle size and shape. However, the composition does have a large impact on the spectral energy distribution (SED). In general, the size and shape distribution of the grains is well constrained by the polarimetric measurements, while the composition is much better constrained from the spectral features seen in the infrared.

In many studies it is assumed that the grains are homogeneous spheres. However, although applicable in some particular cases, the homogeneous dust grain model generally produces resonances that are not observed in the optical properties of natural particles \citep{2011arXiv1111.4348M}. \modified{Though one can often remove the sharp nature of these resonances by using a size distribution of spheres, the smoothed resonances still cause effects that are inconsistent with computations or measurements for irregularly shaped particles, this can lead to errors in the interpretation of observations where the dust grains are expected to be irregularly shaped.}
\citet{2005A&A...432..909M} have shown that breaking the perfect shape of a homogeneous sphere even in the simplest way already reduces the impact of these resonances significantly. Therefore, for this study we will use the DHS grain shape model, as suggested by \citet{2005A&A...432..909M}, using an irregularity parameter $f_\mathrm{max}=0.8$.

We take a grain size distribution with minimum grain size 0.005$\,\mu$m and maximum grain size 0.5$\,\mu$m. We take a powerlaw size distribution and vary the index to match the images.

\subsection{Central star}

The effective temperature and luminosity of the central star are a matter of debate \modified{and varies from 6666\,K \citep{2007MNRAS.374..664C}, derived from optical spectroscopy of the stellar spectrum, to 7500\,K \citep{2003MNRAS.343..880G}, derived from fitting the SED}. The most recent detailed analysis of the optical spectrum \citep{2007BaltA..16..191K} results in an effective temperature of 7250\,K, \modified{which is at the high end of the temperatures found in the literature.} In the same paper these authors also claim that the luminosity of the star is significantly higher than previously assumed, and from this the derived distance must be larger as well. In order to arrive at this claim \citet{2007BaltA..16..191K} assume a relatively high initial mass of the central star ($0.9\,M_{\sun}$), which is at the upper most end of the mass range derived by \citet{2006A&A...450..701S}. Furthermore, \citet{2007BaltA..16..191K} argue that the large reddening of the star points to a large distance, while we will show below that a large fraction of this reddening is circumstellar rather than interstellar. Therefore, we will assume the usual distance of 1.2\,kpc. The luminosity of the star was varied in order to fit the SED. By varying the effective temperature of the star we find that indeed the best fit is obtained by using a Kuruzc model with effective temperature 7250\,K, log$\,g=0.5$, and a luminosity of $3000\,L_{\sun}$.

\subsection{Radiative transfer}

For the radiative transfer we use the highly flexible MCMax code \citep{2009A&A...497..155M}. This code is based on the Monte Carlo radiative transfer method by \citet{2001ApJ...554..615B} and takes into account full anisotropic scattering and polarization. The code has been extensively benchmarked against other codes \citep{2009A&A...498..967P}. The polarimetric images are produced by integrating the formal solution of radiative transfer after the local radiation field at each location in the dust shell is found. Further details can be found in \citet{2009A&A...497..155M, 2011arXiv1111.4348M}.

\subsection{Instrument simulation}

An important part of comparing the resulting model images to the observed images is a proper simulation of instrumental and data reduction effects. We use a detailed simulation of seeing, noise and telescope effects to reduce our high resolution images to the spatial scale and noise level of the observations. The model images are convolved with a PSF matching the seeing conditions at the time of observing. Also, photon noise was added matching the computed fluxes. After this we apply the data reduction pipeline to the simulated observations in exactly the same way as was done for the ExPo observations. The instrument simulator is described in more detail by \citet{2011arXiv1111.4348M}.

\section{Resulting best fit model}

Since we have a large, multi wavelength observational data set, and all observations are of different types, it is very hard to define the best fit in a proper, statistical sense. For this reason we have chosen \modified{to start off by using a by-eye fitting procedure}, where we vary the parameters of the model until we find a solution that best reproduces the observations. This method does not give robust estimates of the uncertainties of the derived parameters, and might be sensitive to local minima in parameter space. The advantage however is that we can judge the goodness of fit for certain datasets and in this way provide relatively robust constraints in certain wavelength intervals. For example, we conclude that we cannot reproduce the mid infrared images properly. We will come back to this issue later in the paper and discuss the implications for the model.

We start from the model by \citet{2002ApJ...571..936M} and varied the parameters for the dust grain size distribution from there until we get a correct match for the polarimetric images. We also checked if we could get a better match by changing the density structure parameters, but find that, \modified{by eye,} a good fit was obtained by taking the parameters A, B, C, E, and F from \citet{2002ApJ...571..936M}. We take the parameter D to be equal to E, which gives a slightly different density structure than obtained by \citet{2002ApJ...571..936M}. If we compute the mass loss history using the value of D given by \citet{2002ApJ...571..936M}, which is D=1, we get a strange dip in the mass loss rate right before the high mass loss phase sets in. Putting D=E, like we did, avoids this problem (see also section~\ref{sec:evolution}). Since the region affected by this is right behind the optically thick torus, it makes almost negligible difference for fitting the SED and the ExPo images.

\modified{In order to more robustly fine-tune the parameters of the model, the second step is to use a genetic fitting algorithm to find the best fit model to the SED, ISO spectrum and ExPo images. A genetic fitting algorithm uses the concept of evolution to find an optimum solution. The procedure starts out with a population of random models. The parameters of the models in this population are treated as their 'genes'. The best models of this population are given the highest change to reproduce their genes (parameter values) into the next generation (survival of the fittest). We use a population of 24 models and evolve this through 100 generations. By studying the resulting 2400 models, we can also get a better feeling which parameters are constrained by the observations and which are more degenerate. For more details on genetic fitting algorithms see \citet{1995ApJS..101..309C}. In order to run this procedure one needs to define what is a good fit. With only a single observational data set this is simply the reduced $\chi^2$. However, since we deal with different datasets which all have different uncertainties and what one would call a good fit is different for different datasets (images are usually harder to reproduce exactly than an SED). Therefore, we have to define the weight we give to each dataset (images and SED). We checked that taking different values for these weights does not make a significant different to the best fit obtained. We fit the radial and azimuthal profiles of the polarized intensity images. The best fit we find does deviate slightly from the fit we get by eye, but all important parameters are very similar. The parameters of this final best fit model are presented in the next section.}

\begin{figure}[!tb]
\centerline{\resizebox{\hsize}{!}{\includegraphics{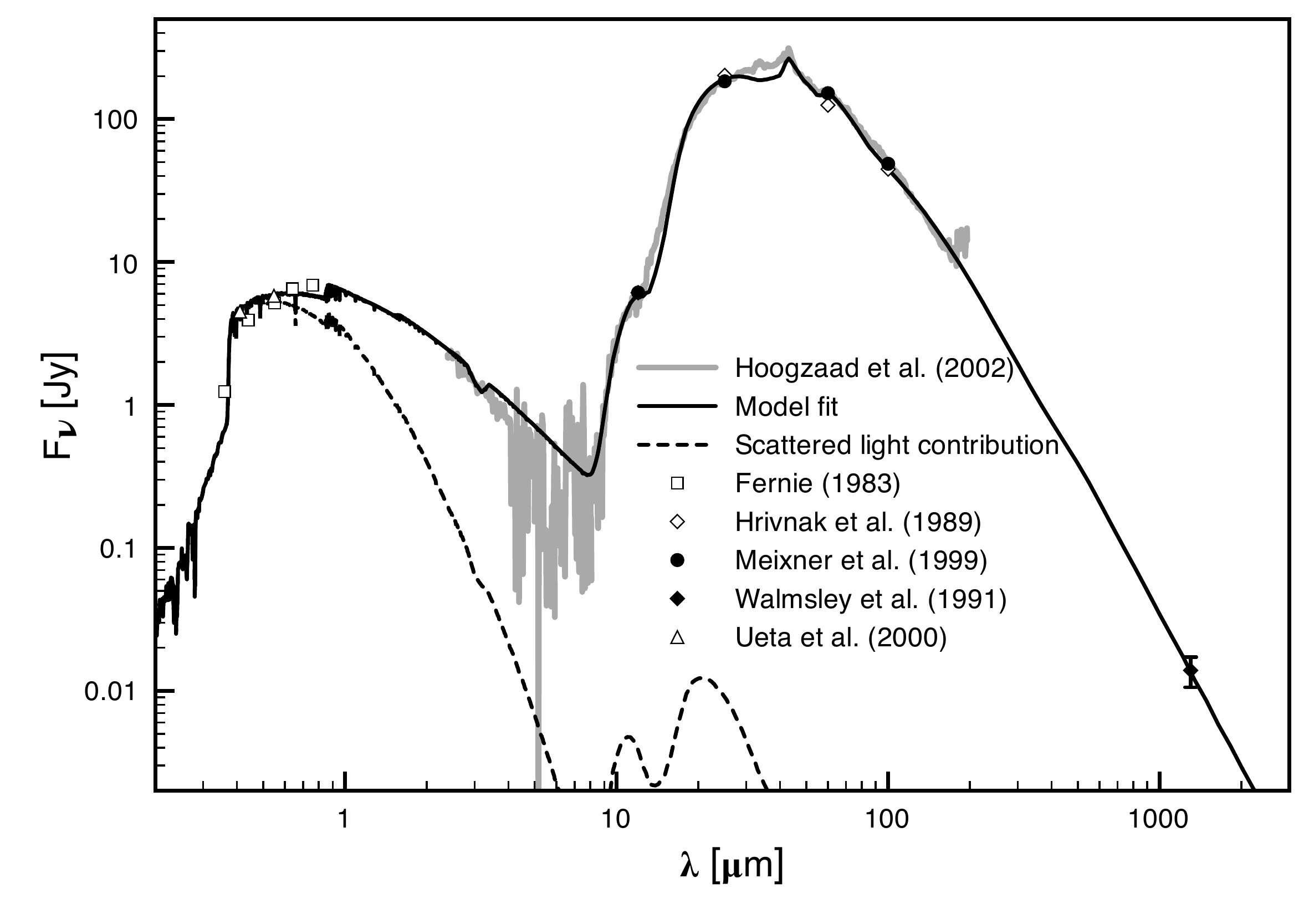}}}
\caption{The spectral energy distribution together with the model fit. ISO spectrum from \citet{2002A&A...389..547H}. Photometry taken from \citet{1983ApJS...52....7F, 1989ApJ...346..265H, 1991A&A...248..555W, 1999ApJS..122..221M, 2000ApJ...528..861U}.}
\label{fig:SED}
\end{figure}

The final model was computed such that it simultaneously reproduces the overall SED and the ExPo polarization images. Also we aim to be roughly consistent with the near infrared imaging polarimetry conducted by \citet{2001MNRAS.322..321G} and the mid infrared intensity images by \citet{2003MNRAS.343..880G}. The resulting SED is shown in Fig.~\ref{fig:SED}, and the simulated images from the model are shown in the lower panels of Fig.~\ref{fig:images}. Although we cannot do absolute photometry, we can compare \modified{the degree of polarization to some extend. We have to be careful since the computed degree of polarization is basically the polarized intensity, caused by scattering off dust grains, divided by the total intensity, which is mostly the tail of the PSF from the central star. Also, the low flux regions in the ExPo images are dominated by noise and unpolarized background emission. Therefore, the computed degree of polarization is influenced by seeing and instrumental effects and cannot be used directly. Since our instrument simulator does take most of these effects into account, we can do a rough comparison. By doing this we find that we systematically over predict the degree of polarization with our model by a factor 3 to 5. This is most likely due to the relatively high degree of polarization of our model particles. Particles with more natural shapes and surface structure usually show a degree of polarization in the range 10-20\,\%, while our model particles have $\sim50\,$\%. We conclude that likely small scale irregularity of the dust grains is the cause for the low degree of polarization observed (see also Section~\ref{sec:intP} for more detailed discussion).}

A notable difference between the model images and the observed images is the lack of elongation at longer wavelengths. We find that, for the model, when the torus becomes optically thin the image becomes roughly circular, while the observations still display an elongated structure. This could indicate that the torus actually starts a bit closer to the star than the bipolar outflow indicating a difference in outflow velocity; the polar outflow being faster by a small fraction. We consider it beyond the scope of our parameterized model to explore this further.

In the model SED we see a rather strong $3\,\mu$m absorption feature caused by the water ice in the torus. This feature is not seen in the ISO spectrum. According to \citet{2002A&A...389..547H} this is caused by an anomaly in the measured spectrum.

\subsection{Derived parameters}

The best fit model parameters are given in Table~\ref{tab:parameters}. We find that the best fit to the SED and observed images is obtained for a nearly edge on geometry, where we look through the torus. \modified{This is consistent with previous modeling of infrared images \citep{2002ApJ...571..936M, 2003MNRAS.343..880G}. The resulting parameters for the shell structure are slightly different than previously reported. This is caused by the additional constraints from the polarimetric images. Although a reasonable fit can be obtained by using the density distribution parameters from \citet{2002ApJ...571..936M}, a better fit to the images is obtained by using the parameters in Table~\ref{tab:parameters}. The values from \citet{2002ApJ...571..936M} are indicated in brackets for the density distribution parameters. Especially the D and E parameters are much higher. This is to create the sharp outer edge of the ring detected in our polarized intensity images. The outer radius of the shell is not constrained by the observations as long as it is large enough to reproduce the far infrared emission.}

\modified{The powerlaw of the size distribution is well constrained.} We have to vary the grain properties in order to have significant optical depth through the equator. Since the absorption optical depth is well constrained by the total amount of reprocessed energy in the infrared, we have to vary the albedo to create additional extinction through scattering. We find that we need grains with a relatively high albedo to get sufficient optical depth at $0.58\,\mu$m to reproduce the observed morphology.

\begin{table}[!tbp]
\caption{\modified{Values for the best fit model parameters}}
\begin{center}
\begin{tabular}{llc}
\hline
\hline
Parameter				&	Symbol		&	Value \\
\hline
Fixed parameters\\
\hline
Effective temperature	&	$T_\mathrm{eff}$	&	$7250\,$K \\
Radius of the star		&	$R_\star$			&	$35\,R_{\sun}$ \\
Luminosity of the star		&	$L_\star$			&	$3000\,L_{\sun}$\\
Distance to the star		&	$D$				&	$1200\,$pc \\
Inclination angle		&	$i$				&	$80^\circ$\\
Minimum grain size		&	$a_{min}$			&	$0.005\,\mu$m\\
Maximum grain size		&	$a_{max}$		&	$0.5\,\mu$m\\
Irregularity parameter	&	$f_\mathrm{max}$	&	$0.8$ \\
\hline
Fit parameters\\
\hline
Inner radius			&	$R_{min}$			&	$904\,$AU \\
Superwind radius		&	$R_{sw}$			&	$970\,$AU \\
Outer radius			&	$R_{out}$			&	$116500\,$AU \\
Dust density at inner edge	&	$\rho_0$			&	$6.7\cdot10^{-21}\,$g/cm$^3$\\
Size distribution powerlaw	&	$p$				&	$3.7$\\
\hline
Density distribution 		&	$A$				&	12.2\quad(8)\\
parameters			&	$B$				&	2.1\quad(2)\\
(in parenthesis are the values		&	$C$				&	2.9\quad(2)\\
from \citet{2002ApJ...571..936M})	&	$D$				&	6.4\quad(1)\\
					&	$E$				&	6.4\quad(3)\\
					&	$F$				&	1.0\quad(1)\\
\hline
\end{tabular}
\end{center}
\label{tab:parameters}
\end{table}

The model we obtain is not unique. Many parameters in our model are to some degree degenerate. However, there are certain things we can constrain with high certainty.

\textit{Optical depth through the equator:} From the ExPo images we can constrain the optical depth through the equator quite accurately. We find that the models can only reproduce the equatorial brightness gap seen in the Na image when the total (i.e. scattering + absorption), optical depth is $\tau\sim 3$.

\textit{Albedo of the grains:} Since we know the total energy reprocessed by the dust grains in the infrared, we actually have good constraints on the optical depth for absorption as well. The reprocessed energy in the infrared should be equal to the amount of energy absorbed in the visual. Together with the total optical depth, this gives us the single scattering albedo of the grains, which in our model is $\omega=C_\mathrm{sca}/C_\mathrm{ext}\sim0.84$ at $\lambda=0.58\,\mu$m. This is actually quite a high value for small cosmic dust grains. Given the assumptions of the composition and shape distribution of the particles that we take, the albedo of the grains accurately constrains the size distribution.

\textit{Inner radius of the outflow:} We directly image the inner radius of the torus. Also from previously published images, the inner radius of the torus is well constrained. We find it to be $0.8''$ which translates to 960\,AU at the assumed distance. This inner radius scales linearly with the distance to the source. \modified{Note that the best fit model puts the inner radius slightly more inwards (904\,AU). However this difference is within the errors caused by our limited spatial resolution.}

\subsection{Degree of polarization}
\label{sec:intP}

\begin{figure}[!tb]
\centerline{\resizebox{\hsize}{!}{\includegraphics{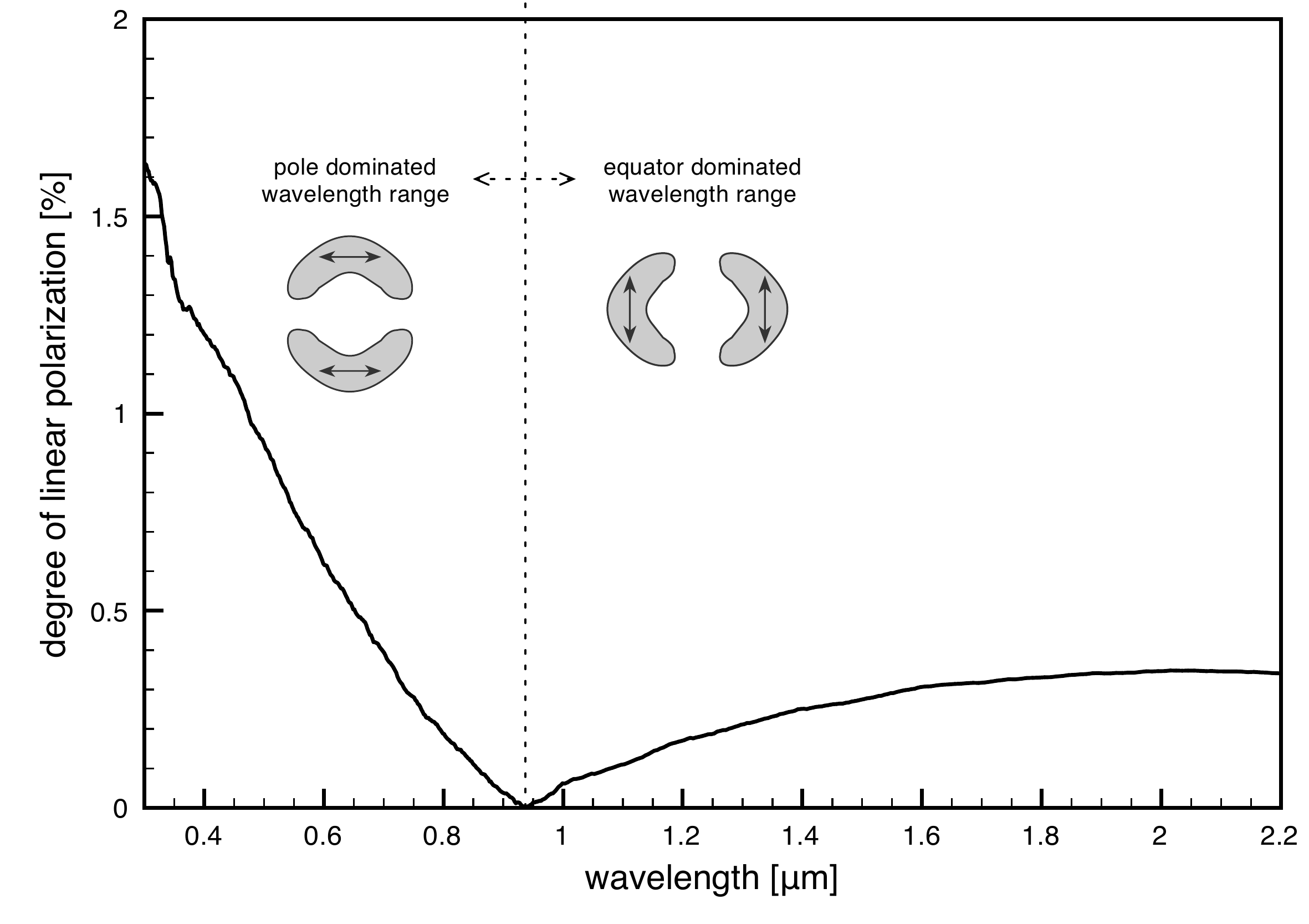}}}
\caption{The total integrated degree of polarization as observed by \citet{2005AJ....129.2451P} together with the computed values from the best fit model. The arrows in the cartoon drawings left and right of the dashed line give the orientation of the polarization vector in the model computations. The vertical dashed line indicates the zero integrated polarization wavelength.}
\label{fig:integratedP}
\end{figure}

\begin{figure}[!tb]
\centerline{\resizebox{\hsize}{!}{\includegraphics{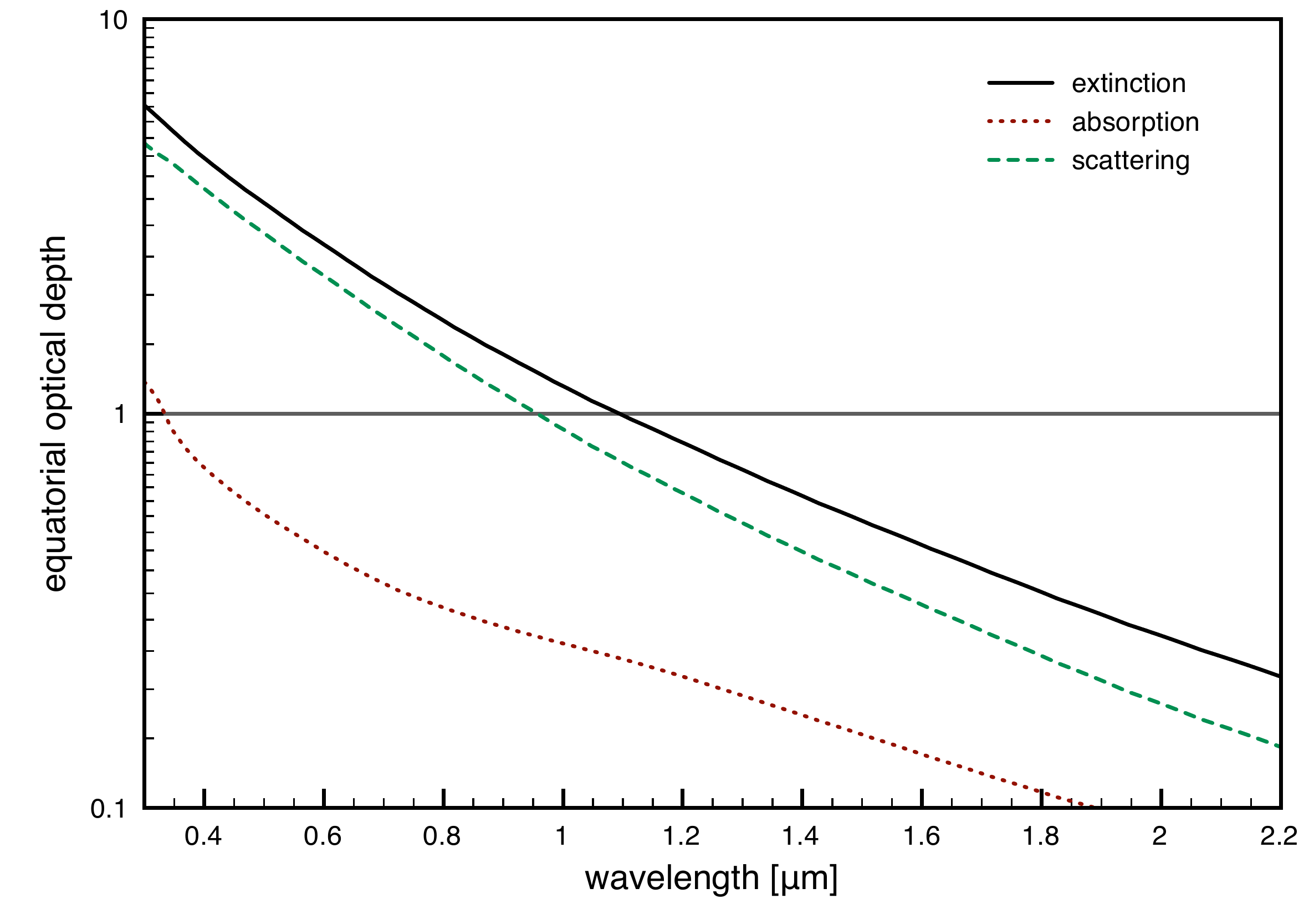}}}
\caption{The equatorial optical depth for scattering, absorption, and total extinction through the torus as a function of wavelength.}
\label{fig:tau}
\end{figure}

\citet{2005AJ....129.2451P} have measured the spatially integrated polarization of HD~161796 to be a bit less than 1\% at optical wavelengths. The orientation of the polarization vectors was found to be roughly east-west (position angle of $\sim110^\circ$). \modified{We cannot compare these observations directly to the model calculations since the observations are likely heavily polluted by interstellar polarization.} Since ExPo is mounted in Nasmyth focus, we have to deal with a relatively high degree of instrumental polarization. This is removed by assuming that the central star is unpolarized, a method which is great for enhancing spatial variations in the polarized intensity. However, it implies that we can not do absolute polarimetry using the ExPo images, and thus cannot compare our values with those from \citet{2005AJ....129.2451P}. \modified{In Fig.~\ref{fig:integratedP} we show the values for the integrated polarization predicted by our best fit model. We predict from the model that the angle of polarization will flip 90 degrees when going to the near infrared.} This is because at the short wavelengths scattered light from the material ejected at the poles dominates, while at the longer wavelengths the scattered light from the torus takes over (see also the simulated near infrared images in Fig.~\ref{fig:NIR images}). The turnover point in our model is approximately when the optical depth through the torus equals unity (see Fig.~\ref{fig:tau}). This result is consistent with the measured polarization angle in the optical by \citet{2005AJ....129.2451P} and with the polarization images in the near infrared from \citet{2001MNRAS.322..321G}.

\begin{figure}[!tb]
\centerline{\resizebox{\hsize}{!}{\includegraphics{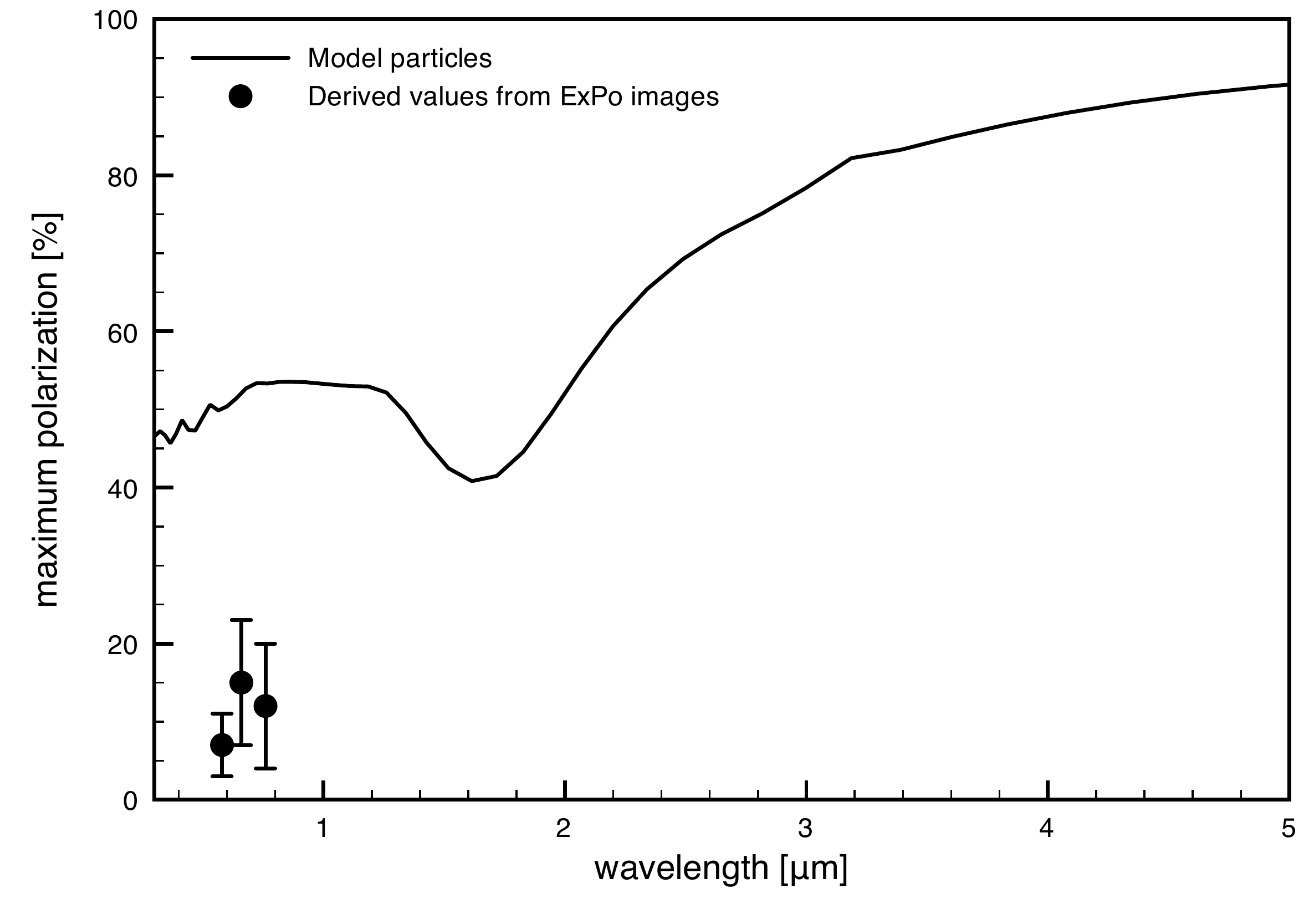}}}
\caption{\modified{The maximum degree of polarization of single scattering as a function of wavelength. The maximum value is usually around a scattering angle of 90 degrees. The values derived from the observations have been found by scaling the model images to match the observations. The scaling factor directly gives the scaling that has to be applied to the model particles.}}
\label{fig:Pmax}
\end{figure}

There are two solutions to the high degree of polarization we predict with respect to the observations. Lowering the scattering cross section of the grains, and lowering the degree of polarization of the grains. Lowering the scattering cross section is difficult because it implies lowering the overall optical depth. This means that the torus has a lower optical depth, already at $0.58\,\mu$m which is inconsistent with the decreased equatorial brightness seen in the ExPo images. Lowering the degree of polarization seems more likely. In the model grains we use, the degree of polarization is roughly constant over the visible part of the spectrum. \modified{The observations seem to indicate that the degree of polarization at wavelengths shorter than $0.7\,\mu$m is significantly lower than we predict, up to almost a factor of five at $0.58\,\mu$m. This can be effects of surface roughness or internal irregularities of the grains. The grains we model have a grey polarization color, i.e. the degree of polarization is unchanged when going from 580 to 760\,nm wavelength. In reality the color of polarization is likely red, i.e. the degree of polarization is lower at shorter wavelengths \citep[like for example is the case for cometary dust][]{2010EP&S...62...17K}. In Fig.~\ref{fig:Pmax} we show the maximum degree of polarization, usually occurring around 90 degrees scattering, as a function of wavelength. We show both the values for the model particles as well as the values inferred from the mismatch between model and ExPo images. These observationally derived intrinsic degrees of polarization of the dust particles are derived by equating the model images and the observations. These values have large uncertainties, but seem to be in line with the steady decline with decreasing wavelength as seen wavelengths longwards of 1.6\,$\mu$m. However, the model curve increases with decreasing wavelength shortwards of 1.6\,$\mu$m. This is likely due to an increased contribution from surface scattering at short wavelengths. Surface scattering off a smooth surface causes high degrees of polarization at the Brewster angle. This is not observed for scattering off rough surfaces. We therefore tentatively attribute the difference between model and observed degree of polarization in the optical to surface roughness of the grains. Interestingly, at wavelengths longwards of 1.6\,$\mu$m the degree of polarization predicted by our model is in line with the observations (see Section~\ref{sec:NIR}). This is what one would expect from this argumentation, since at long wavelengths surface effects are negligable.} Detailed modeling of this is computationally challenging and beyond the scope of this paper.

\subsection{HST image}

\modified{\cite{2000ApJ...528..861U} present an image taken with the Hubble Space Telescope (HST) of HD~161796. This image displays a strong centrally peaked object with no clearly resolved outer structure. For an optically thin shell, one would expect to see the limb brightened inner edge of the shell clearly resolved. However, since the shell is optically thick at visible wavelengths, this effects is not observed. The polarized flux images enhance the outer edges of the shell in the image, since the forward scattered light is unpolarized, and thus filtered out. The projected edges of the shell scatter at approximately 90 degrees scattering angle, and are thus highly polarized. This makes detecting the edge of the shell possible in polarized flux (our ExPo images), but not show up in scattered light images of total intensity (HST image). Indeed, a simulated HST observation of our model shows a morphology consistent with the image presented in \citet{2000ApJ...528..861U} (see also the middle panel of Fig.~\ref{fig:timeshift}, which is discussed in section~\ref{sec:evolution}).}

\subsection{Near infrared images}
\label{sec:NIR}

\begin{figure}[!tb]
\centerline{\resizebox{0.8\hsize}{!}{\includegraphics{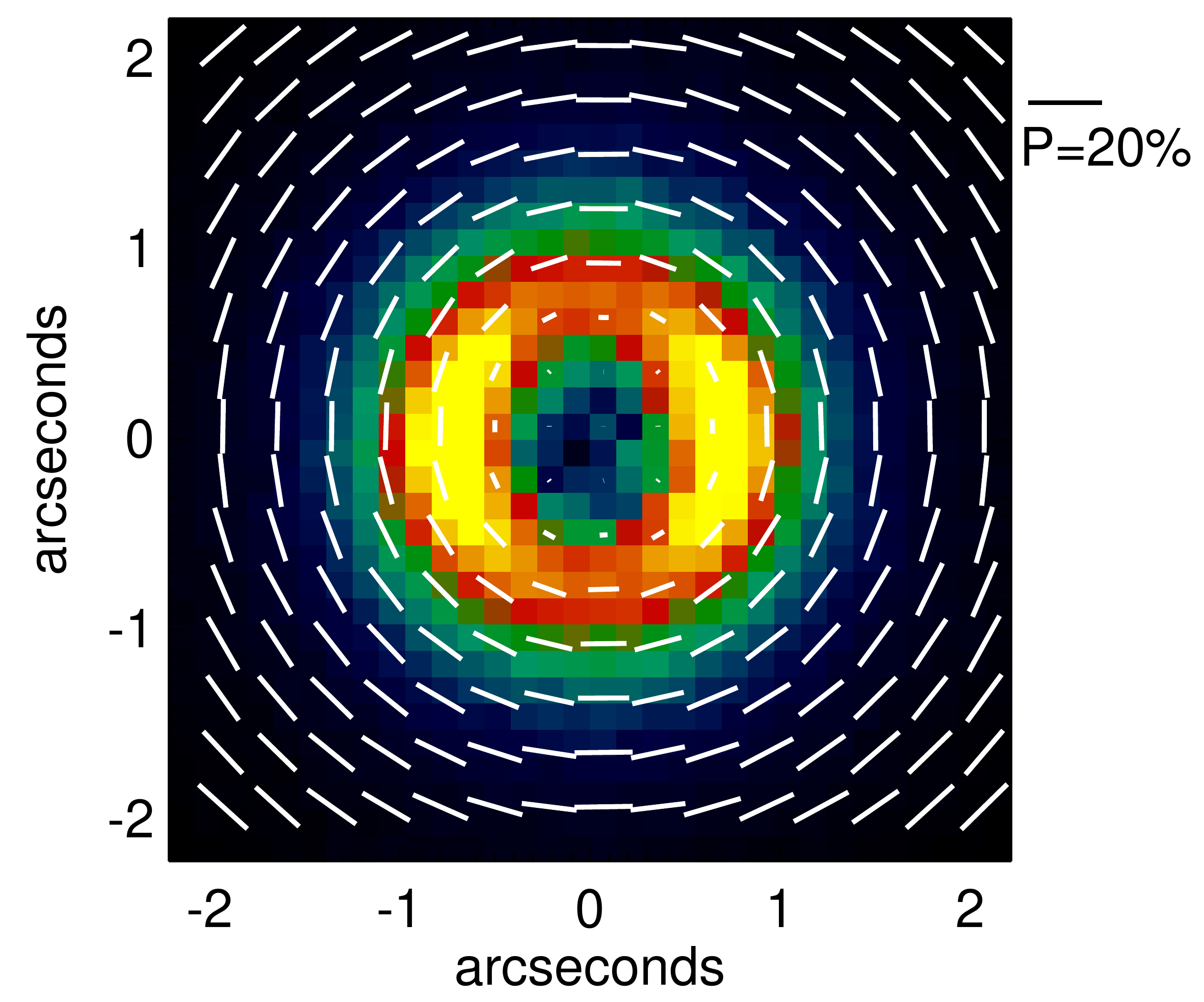}}}
\centerline{\resizebox{0.8\hsize}{!}{\includegraphics{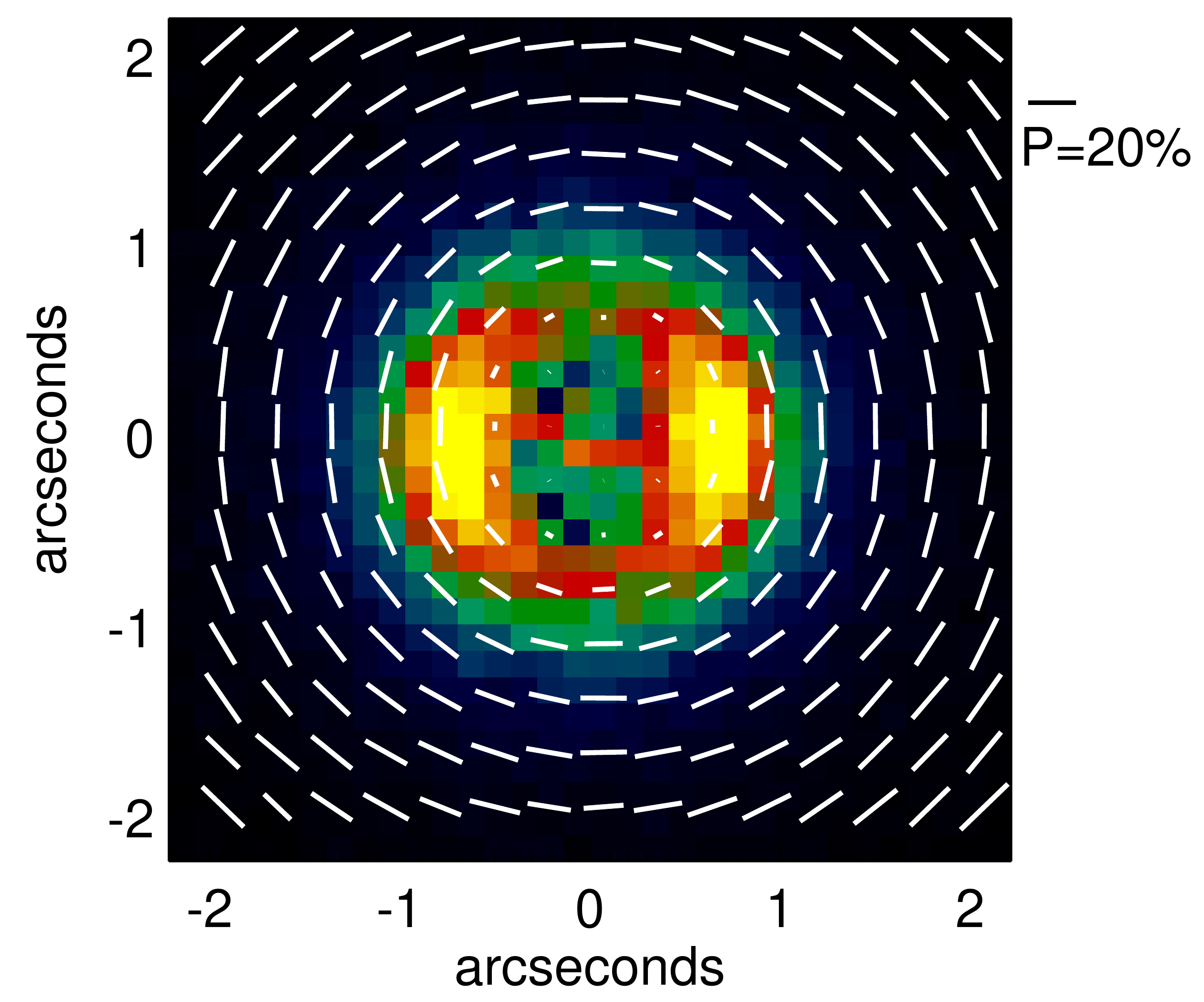}}}
\caption{\modified{Model images of the polarized flux in the J (upper panel) and K (lower panel) bands. To be compared with observations from \citet{2001MNRAS.322..321G}. The colormap is the same as Fig.~\ref{fig:images}.}}
\label{fig:NIR images}
\end{figure}

\citet{2001MNRAS.322..321G} have obtained polarimetric images of HD~161796 in the near infrared using the 3.8\,m United Kingdom Infrared Telescope (UKIRT) on Mauna Kea. The obtained images are in the J and K band and they find that there are two bright blobs located east and west of the central star. This structure is more pronounced in the longer wavelength K-band image than in the J-band image. \modified{They report degrees of polarization up to $\sim20\,$\%.} We constructed model images at these wavelength, convolved them with a psf from a 3.8\,m mirror and smeared them with a seeing of $0.6''$, similar to the resolution they claim. Also we added some photon noise to simulate the observations. The resulting images are shown in Fig.~\ref{fig:NIR images}. Comparing these images to Fig.~3 in \citet{2001MNRAS.322..321G} we see that we have a very nice match. Indeed, at these wavelengths the equatorial structure dominates the scattered light. \modified{Also, the modeled degree of polarization at these wavelengths matches the observations.}

\subsection{Mid infrared images}

\begin{figure}[!tb]
\centerline{\resizebox{0.8\hsize}{!}{\includegraphics{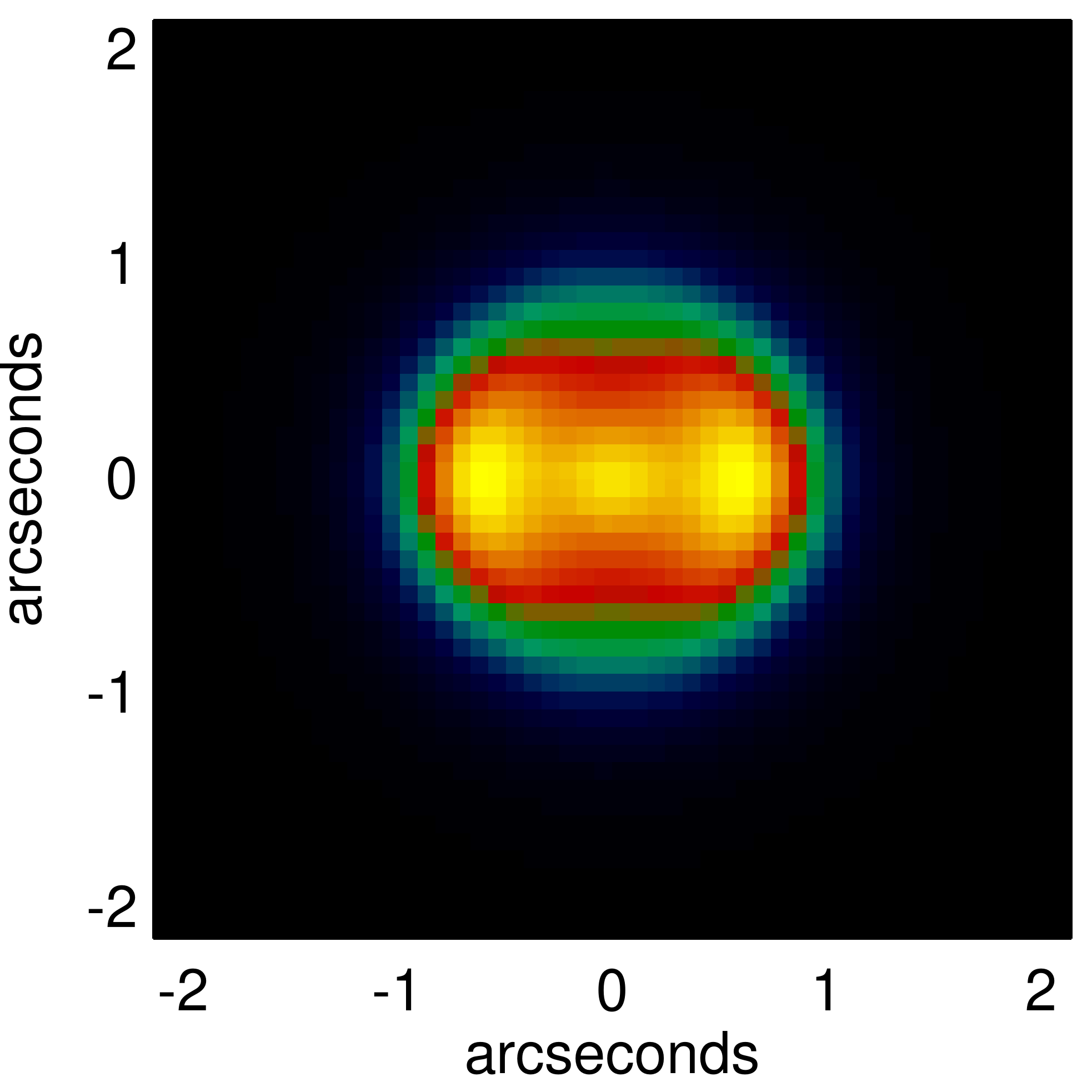}}}
\caption{Model image at 12.5\,$\mu$m.}
\label{fig:IR images}
\end{figure}

Mid infrared images of HD~161796 have been obtained by \citet{1994ApJ...423L.135S} and later by \citet{2003MNRAS.343..880G}. Around $10\,\mu$m these images display two bright blobs east and west of the central star, while the central star itself is not prominent in the image. A reconstructed image at $12.5\mu$m convolved with a psf of $0.5''$, similar to the observations, does not fully reproduce this behavior. We find that we still see a contribution from the central star (see Fig.~\ref{fig:IR images}). We have put significant effort into solving this problem, but without success. Increasing the contrast between the equatorial structure and the star requires additional absorption in the infrared. However, we find that when we do this the total reprocessed luminosity goes up, so the SED does not fit anymore. A way to resolve this would be reducing the optical absorption while keeping the infrared extinction the same. We can do this by removing the iron from the silicate grains. However, we find that in this case the silicates become too cold and we have to put them closer in to fit the SED. However, this is in contrast with the images that actually constrain the inner radius of the dust grains very well. In addition, increasing the optical depth at $10\,\mu$m causes the bright blobs to be less distinctive, and the fit to the SED becomes much worse. We conclude that at this moment we do not have a solution for this apparent discrepancy.

We note here that the total infrared flux reported by \cite{2003MNRAS.343..880G} from the images is significantly higher than those obtained from the ISO spectrum or the fluxes measured by \citet{1989ApJ...346..265H} and \citet{1999ApJS..122..221M}. If we adjust the model to match the flux derived by \cite{2003MNRAS.343..880G} we do find a solution where the bright torus dominates over the light from the central resolution element. However, the higher flux levels reported there cannot be explained by the mild variations of the central source. Also, we note that we did not succeed in reproducing the fit to the infrared images obtained by \cite{2003MNRAS.343..880G} or \citet{2002ApJ...571..936M}. Unfortunately, the model images in these papers are only presented in contour plots and not in detailed grey-scale plots, which makes a detailed comparison difficult. At this point we must conclude this is still an unsolved issue.

%
%
%

\section{Mass loss history}
\label{sec:evolution}

\begin{figure}[!tb]
\centerline{\resizebox{\hsize}{!}{\includegraphics{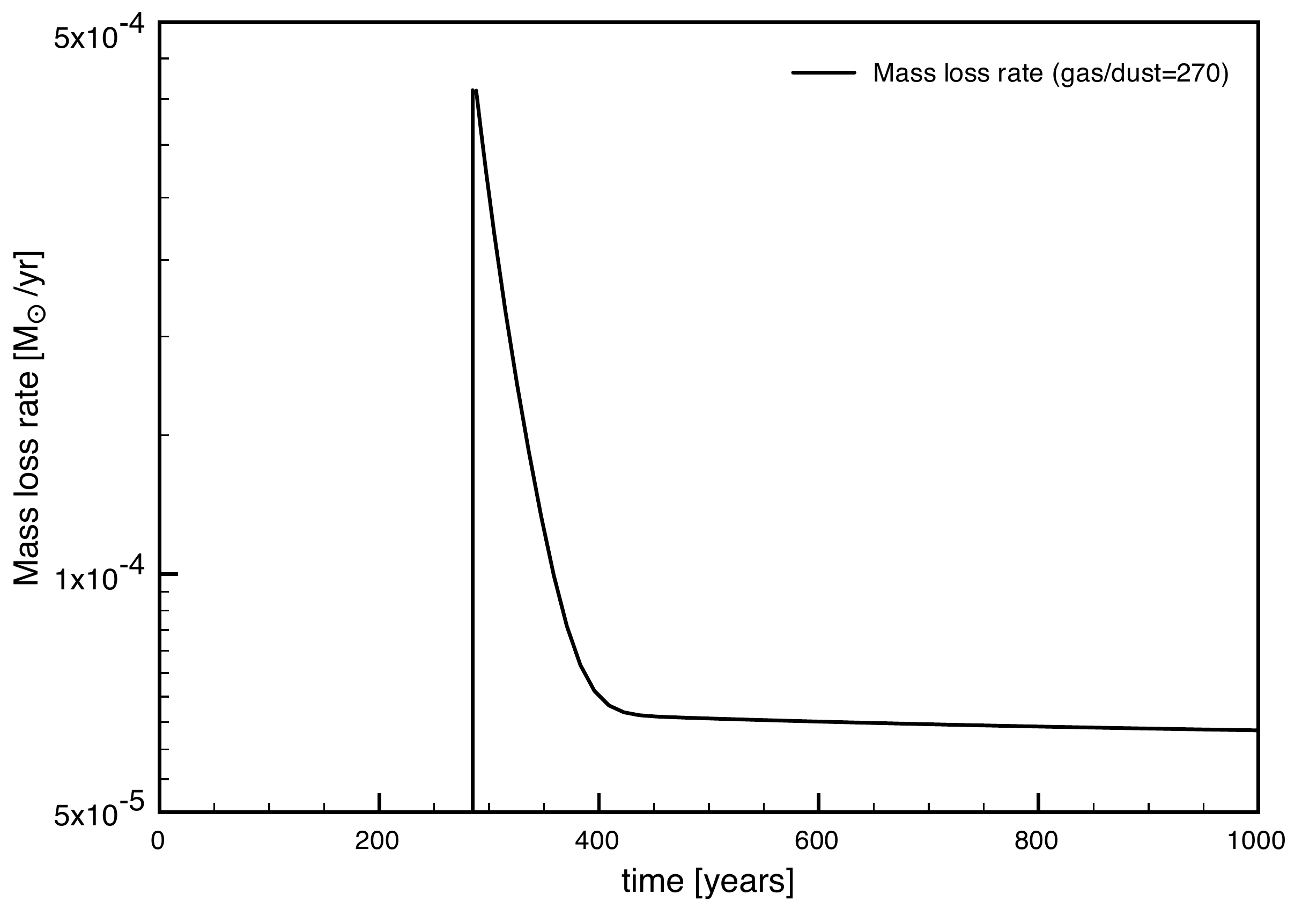}}}
\caption{Mass loss history of HD~161796 as derived from the density of the model. The total mass loss rate is plotted using a gas/dust ratio of 270 \citep{2002A&A...389..547H}.}
\label{fig:masslosshistory}
\end{figure}

\modified{We find a slightly smaller inner radius than \citet{2002A&A...389..547H}, so we find that the mass loss of HD~161796 stoped around 285 years ago in stead of 430 years as computed by \citet{2002A&A...389..547H}.} When we assume a constant outflow velocity (we take 15\,km/s), we can compute the mass loss history of the object from our derived density structure. This is shown in Fig.~\ref{fig:masslosshistory}.

\begin{figure*}[!tb]
\centerline{\resizebox{\hsize}{!}{\includegraphics{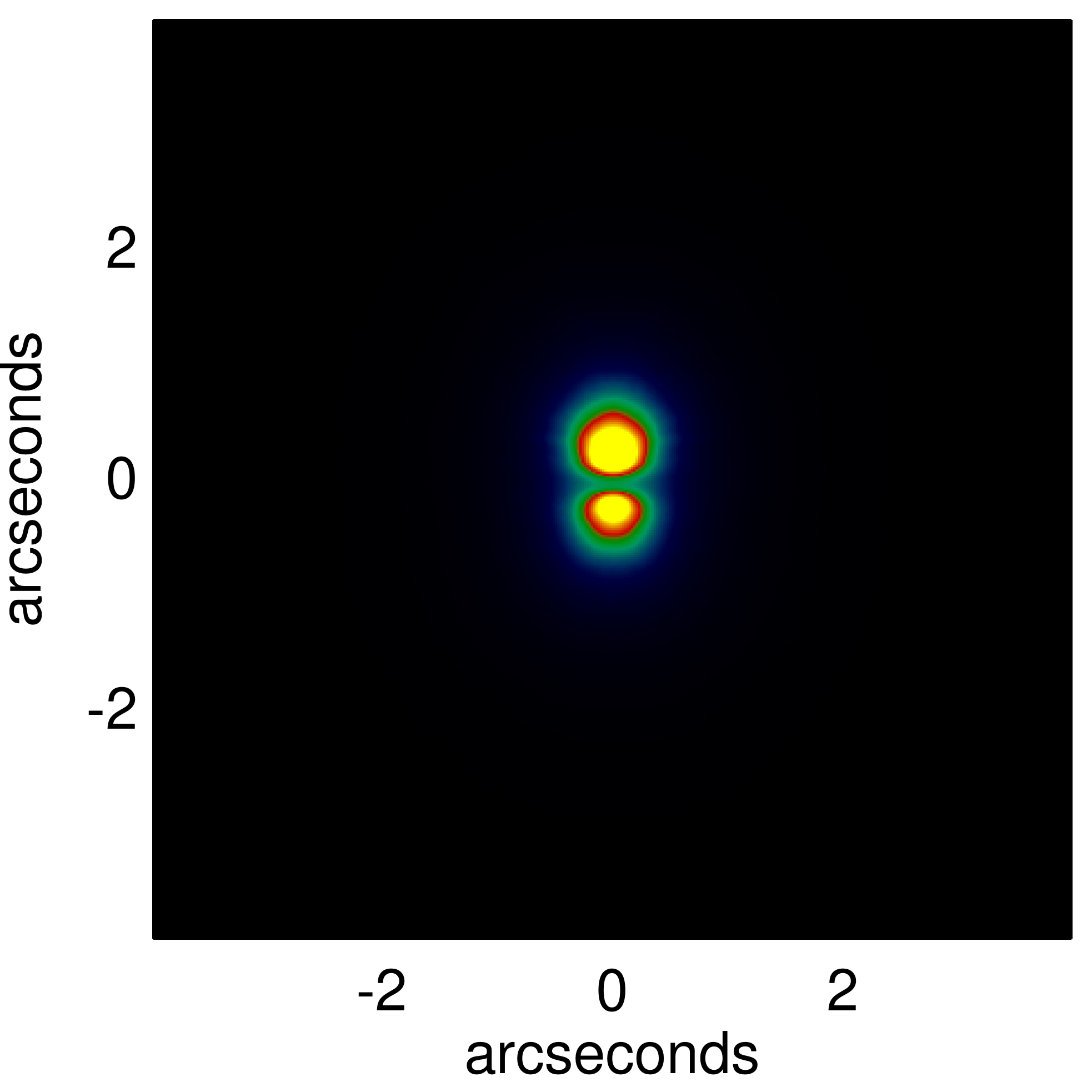}{\includegraphics{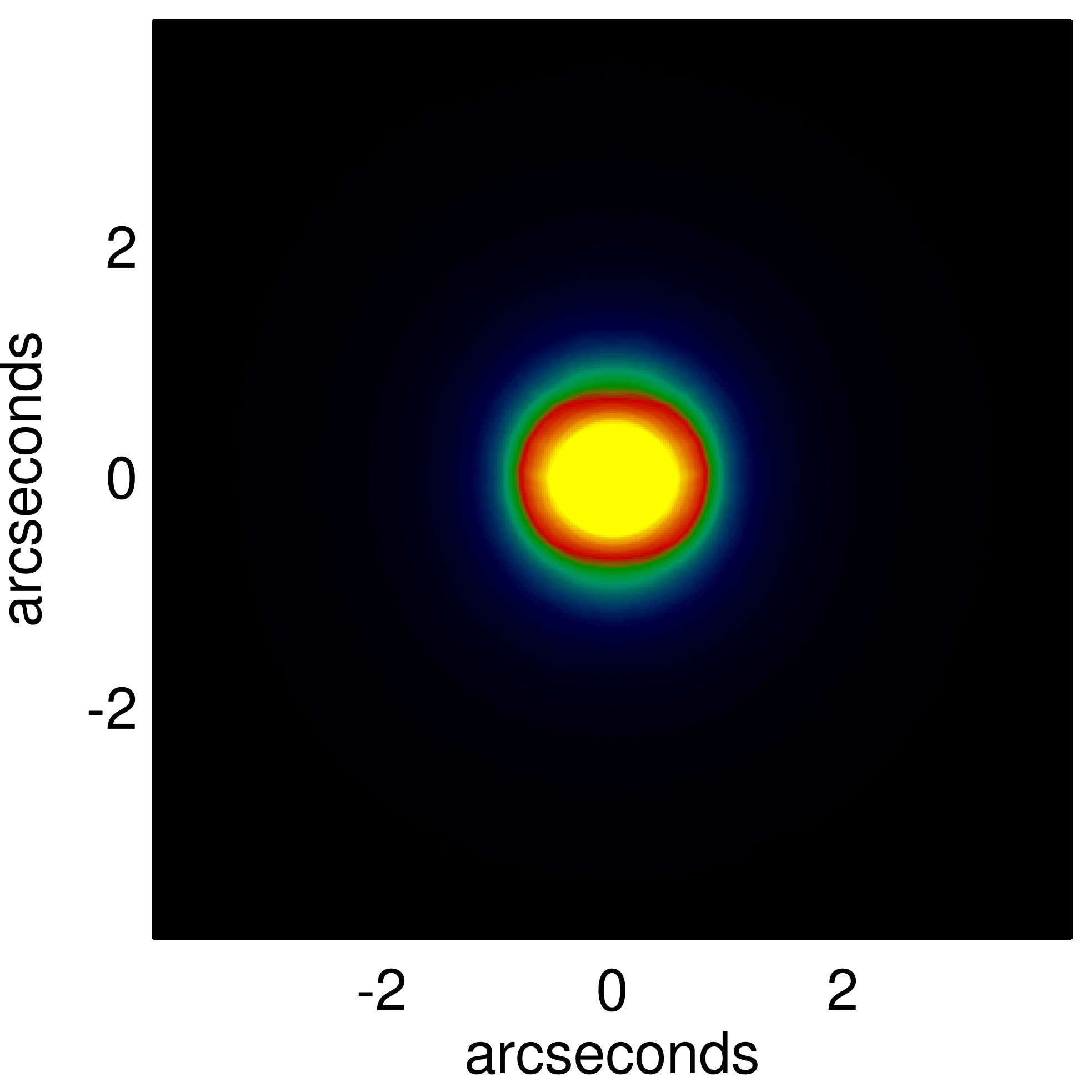}}{\includegraphics{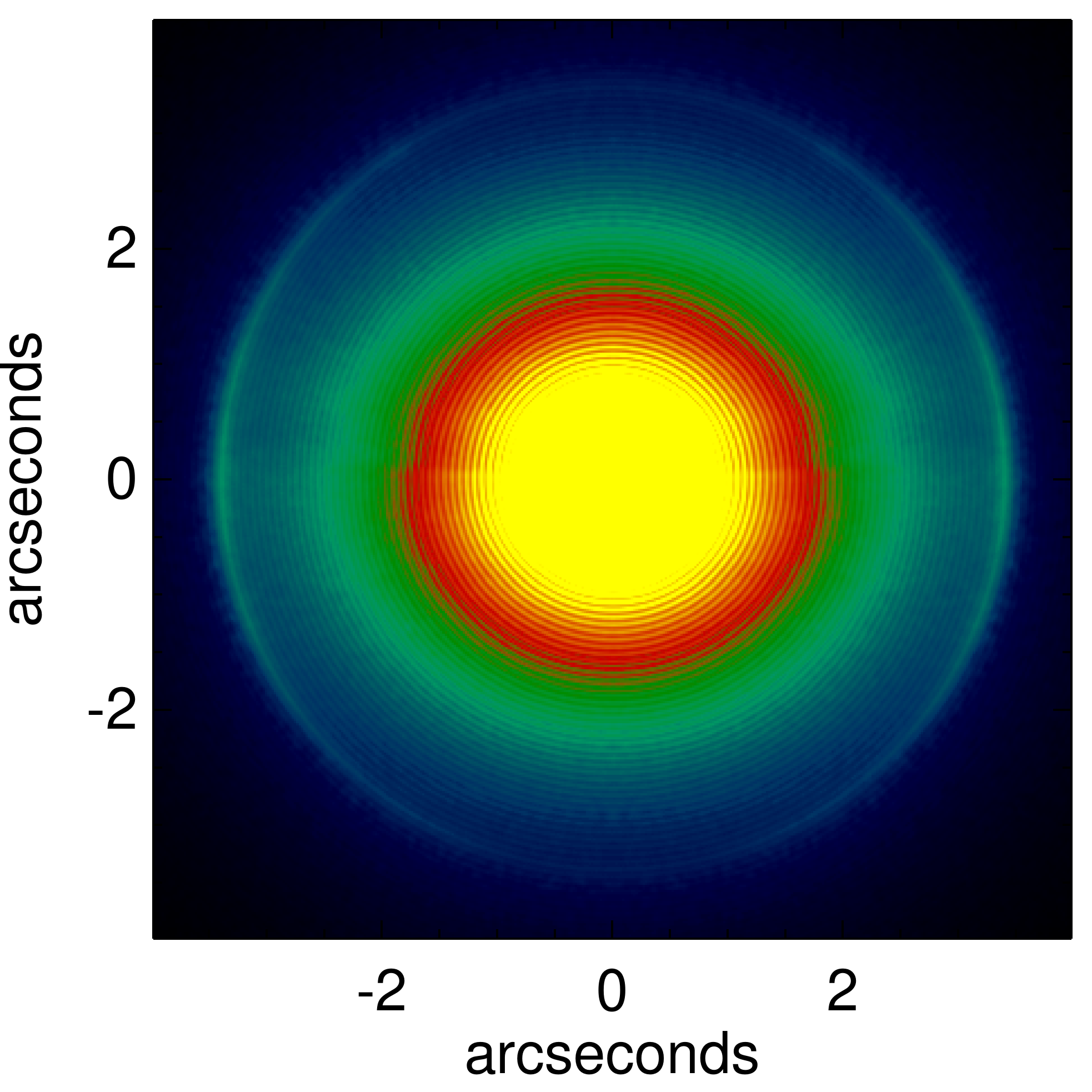}}}}
\caption{Model images of the total intensity as a function of time, assuming a constant outflow velocity of 15 km/s. The model images are copmputed at $\lambda=0.75\,\mu$m, and convolved with a point spread function representative for an HST type telescope. The left image is computed for 250 years ago, the middle image is for the present day, and the rightmost image is computed for 1000 years from now.}
\label{fig:timeshift}
\end{figure*}

It is interesting to see what the derived density structure of the circumstellar shell would have looked like in the past. We can easily compute backwards, assuming a constant outflow velocity of the material, to take snapshots of the nebula at different times. Also, we can predict what the nebula will look like in a thousand years. Model images are shown in Fig.~\ref{fig:timeshift} for $\lambda=0.75\,\mu$m, and convolved with a PSF typical for a HST type telescope. We show here the image from 250 years ago since the image computed for 285 years ago, right after the mass loss stopped, is completely obscured at visible wavelengths. It is clearly seen that in the evolution from 250 years ago, when the star was still a heavily obscured OH/IR type star, to 1000 years from now, the image morphology changes significantly. This is an effect of the changing optical depth through the circumstellar envelope. \modified{In the image close after the mass loss stopped, the optical depth through the shell is still very high. This results in the bipolar structure observed in the leftmost panel of Fig.~\ref{fig:timeshift}. At the present day, the optical depth through the shell is much less, and the central bright component dominates. In the image computed for 1000 years in the future (rightmost panel in Fig.~\ref{fig:timeshift}) the envelope has reached a distance from the central star far enough to be resolved. Also, the optical depth through the shell has decreased such that now the arcs to the east and west dominate in stead of the bipolar structure seen at early times.}

\section{Conclusions}

We have imaged HD~161796 in the optical using imaging polarimetry in various filters. We show that the structure seen varies from dominated by the low density polar outflow at short wavelength to a structure dominated by the high density equatorial regions at long wavelengths. This is easily explained by an equatorial density enhancement which is optically thick at short wavelengths and changes to optically thin at longer wavelengths. We show that the point where this change in geometry takes place for the dust shell around HD\,161796 lies around $1\,\mu$m, where the optical depth through the equator is around unity. \citet{2000ApJ...528..861U} classified the two different appearances of proto-planetary nebulae as SOLE (optically thin torus) and DUPLEX (optically thick torus). We show that for HD~161796 this classification is wavelength dependent showing that indeed these are two different manifestations of the same physical structure.

We present a model for the circumstellar shell which is consistent with most observational constraints. We find that in the optical the integrated flux is dominated by scattered light, while the central star is obscured from sight by the flattened, equatorial structure. This implies that the reddening of the central star is predominantly circumstellar. In order to explain our polarimetric images we need to have dust grains present in the torus with a relatively high albedo of $\sim0.84$. We achieve this by using grains with moderate sizes. When the grains are too large, the albedo becomes too high, when they are too small, the albedo becomes too low. Given the assumptions we made for the composition and shape distribution of the grains we can constrain the size distribution accurately to range from $0.005-0.5\,\mu$m with a power law distribution given by $n(a)da\propto a^{-3.7}$.

Our model is consistent with the SED and with the images available from $0.58 - 2.2\,\mu$m. The model is roughly consistent with the observed mid infrared images with the important exception that the central star remains visible in our model, while it is not seen in the observed images. We have put significant effort into solving this problem, without success. This is an issue which should be addressed in future detailed modeling of this object.

\modified{We find that the degree of polarization observed is significantly lower than what is computed using a simplified dust model. Model particles show an unrealistically high degree of polarization caused by scattering off an infinitely smooth surface. Therefore, we attribute the difference between observation and model to surface roughness and small scale irregularities of the dust grains surrounding HD~161796.}

Finally, we find that the central star is significantly hotter than often assumed. This is mainly caused by the fact that the star is obscured almost completely by the equatorial structure. This causes a change of the spectrum of the star which makes it appear cooler than it actually is. This is similar to the recent analysis by \citet{2007BaltA..16..191K}. For this high temperature to be consistent with the observed SED, it is needed to significantly redden the source. \citet{2007BaltA..16..191K} suggest that this reddening is interstellar, requiring a very large distance, while we show that the reddening is circumstellar, so we do not need to increase the distance to the source.

\begin{acknowledgements}
We would like to thank Hans van Winckel for useful discussions on the properties of the central star. H.C.C. acknowledges support from Millenium Science Initiative, Chilean Ministry of Economy, Nucleus P10-022-F. 
\end{acknowledgements}

\end{document}